\documentclass[peerreview,letterpaper, 11pt]{IEEEtran}

\usepackage[utf8]{inputenc} 
\usepackage[T1]{fontenc}
\usepackage{url}
\usepackage{ifthen}
\usepackage{cite}
\usepackage[pdftex]{graphicx}
\graphicspath{ {./figures/} }
\usepackage[cmex10]{amsmath} 
\usepackage{tikz}
\usetikzlibrary{positioning, shapes.geometric}
\usepackage{dansong} 
\usepackage{xcolor}
\usepackage{xspace}
\newcommand{\blue}[1]{#1}
\usepackage[capitalize]{cleveref} 

\newtheorem{cor}{Corollary}
\newtheorem{definition}{Definition}
\newtheorem{example}{Example}
\newtheorem{lemma}{Lemma}
\newtheorem{prop}{Proposition}
\newtheorem{remark}{Remark}
\newtheorem{theorem}{Theorem}

\crefformat{equation}{\textup{(#2#1#3)}}
\Crefformat{equation}{\textup{(#2#1#3)}}

\newcommand{\alphfont}[1]{\mathcal{#1}}
\newcommand{\seqfont}[1]{\mathbf{#1}}
\newcommand{\empirical}[1]{Q\left[#1\right]}

\newcommand{\assumefinitealphabet}{Suppose the alphabets $\xc, \zc, \yc$ are finite.}

\begin{document}
\title{The Performance of Compression-Based Denoisers} 

\author{%
\IEEEauthorblockN{Dan~Song and Ayfer~\"Ozg\"ur and Tsachy~Weissman}

\IEEEauthorblockA{Department of Electrical Engineering\\
                    Stanford University\\
                    Stanford, CA\\
                    Email: \{songdan, aozgur, tsachy\}@stanford.edu}
\thanks{This paper was presented in part at ISIT 2025 \cite{11195706}.}
}


\maketitle


\begin{abstract}
We consider a denoiser that reconstructs a stationary ergodic source by lossily compressing samples of the source observed through a memoryless noisy channel.
Prior work on compression-based denoising has been limited to additive noise channels.
We extend this framework to general discrete memoryless channels by deliberately choosing the distortion measure for the lossy compressor to match the channel conditional distribution.
By bounding the deviation of the empirical joint distribution of the source, observation, and denoiser outputs from satisfying a Markov property, we give an exact \blue{asymptotic} characterization of the loss achieved by such a denoiser.
Consequences of these results are explicitly demonstrated in special cases, including for MSE and Hamming loss.
A comparison is made to an indirect rate-distortion perspective on the problem.
\end{abstract}
\section{Introduction} \label{sec:intro}
    \newcommand{\loss}{\rho}
\newcommand{\X}{X}
\newcommand{\xs}{\seqfont{\X}}
\newcommand{\Z}{Z} 
\newcommand{\zs}{\seqfont{\Z}}
\newcommand{\Y}{Y} 

\newcommand{\thedensity}{p}
\newcommand{\thedist}{P}

\newcommand{\Pzgx}{\thedist_{\Z|\X}}

\begin{figure}
    \centering
    \begin{tikzpicture}[node distance=0.38\linewidth, every node/.style={}]
    
    \node[font=\Large] (X) {$\X^n$};
    \node[font=\Large, right=of X] (Z) {$\Z^n$};
    \node[font=\Large, right=of Z] (Y) {$\Y^n \approx \X^n$};
    
    \draw[->] (X) -- (Z) 
    node[midway, draw, rectangle, inner sep=4pt,fill=white, align=center] 
    {Memoryless Channel\\$\Pzgx$};
    
    \draw[->] (Z) -- (Y) 
    node[midway, draw=blue, fill=white, rectangle, inner sep=4pt, align=center] 
    {``Good'' lossy\\source code\\ for $\rho$ at $D$};

\end{tikzpicture}
\caption{Setting considered in this work. The source $\X^n$ and the channel $\Pzgx$ are fixed and known, and $\rho$ and $D$ are designed so that the reconstruction $\Y^n$ is close to $\X^n$.}
\label{fig:general-setup}
\end{figure}

Consider the setting in \Cref{fig:general-setup}, where $\X^n$ is generated by a stationary ergodic source --- not necessarily i.i.d.--- and observed through a known memoryless channel $\Pzgx$, producing the observations $\Z^n$. In this work, we study the recovery of $\X^n$ by lossily compressing the observations $\Z^n$ into reconstructions $\Y^n$. We define a distortion measure $\loss$ between $\Z^n$ and $\Y^n$ that depends only on the channel $\Pzgx$, and show that when the sequence of noisy observations $\Z_i$ is compressed using a lossy compressor optimized for $\loss$ at a specific distortion level $D$, the resulting reconstructions $\Y^n$ effectively serve as a denoising of the source sequence. 

Our setting can be contrasted with the classical problem of indirect rate-distortion \cite{witsenhausen_indirect_1980,dobrushin_information_1962}, which also has a source $\X^n$ observed through a noisy channel.
In the classical indirect rate distortion setting, one begins with a prescribed distortion measure between $\X^n$ and $\Y^n$, and the lossy compressor is optimized for this particular distortion measure. In contrast, in our framework, no such distortion measure between $\X^n$ and $\Y^n$ is specified a priori; instead, we want the compression to denoise the observations (in the sense of ``inverting'' the impact of the noisy channel) so that the fidelity of the resulting reconstructions can be universally bounded with respect to \emph{any} distortion measure between $\X^n$ and $\Y^n$.

Leveraging the idea that compression inherently removes noise to perform denoising has appeared in the prior literature in various settings \cite{natarajan_filtering_1995, natarajan_occam_1998, rissanen_mdl_2000, donoho_kolmogorov_2002,weissman_empirical_2005}. In \cite{natarajan_filtering_1995} and \cite{natarajan_occam_1998}, Natarajan introduces Occam filters, which apply this idea to remove additive noise from real-valued signals, using lossy compressors operating at a norm distortion equal to the norm of the noise. Upper bounds on the expected norm between $\X^n$ and $\Y^n$ (treated as vectors in $\reals^n$) are given in terms of the operating rate of the compressor and the rate-distortion function for the noise source. This theoretical result is limited in that it depends on the specific properties of the compressor used, and the rate-distortion function of the observations may in general be difficult to evaluate. The Occam filters are shown to perform well empirically, but the upper bounds are loose in practical regimes.

Our work is most closely related to that of \cite{donoho_kolmogorov_2002} and \cite{weissman_empirical_2005}.
In \cite{donoho_kolmogorov_2002}, Donoho proposes using lossy compression with distortion chosen to match the amount of error introduced by the noise to recover samples from the posterior in two cases: a binary source passed through a binary symmetric channel and a Gaussian source passed through an AWGN channel. He gives bounds for the Hamming and squared losses achieved in the two settings, respectively.
In \cite{weissman_empirical_2005} Weissman and Ordentlich
consider the joint empirical distribution of the source and reconstruction sequences of a compressor.
Under some regularity conditions, they show empirical distributions of ``good'' compressors approach the distribution that achieves the infimum in the definition of the rate-distortion function.
They apply this result to the denoising setting where a source is corrupted by i.i.d. additive noise, i.e., $\Z_i = \X_i + N_i$.
They show that choosing a distortion measure $\log p_N(z-y)$ and distortion level $H(N)$ suffices for a lossy compressor to recover the source, in the sense that the reconstruction asymptotically behaves as samples from the posterior distribution.
This characterizes the empirical distribution of $\X^n, \Z^n$ and $\Y^n, \Z^n$ (but not the joint distribution of $\X^n, \Z^n$ and $\Y^n$), and \cite{weissman_empirical_2005} gives a bound on the performance of compression-based denoising by assuming the worst case coupling between the two marginal distributions $\X^n, \Z^n$ and $\Y^n, \Z^n$. 

Our results both generalize and strengthen the findings of prior works, including~\cite{weissman_empirical_2005}, which are limited to additive noise channels. We extend the compression-based denoising framework to arbitrary memoryless channels $\Pzgx$ by identifying a suitable distortion measure that is chosen to match $\Pzgx$. When a compressor is optimized for this distortion measure, it effectively removes the influence of the noisy channel. This same distortion measure was first introduced in~\cite{reshetova_training_2024} as a cost function for optimal transport in the context of training generative models from privatized data, where it similarly serves to mitigate the effects of privatization and enables the model to learn the underlying raw data distribution. Furthermore, we show that under the joint empirical distribution of $\X^n$, $\Y^n$ and $\Z^n$, the noise-free and reconstruction variables are essentially conditionally independent given the noisy observations. Thus good lossy compression under the right noise-induced distortion criterion and level not only results in a "sample from the posterior", but an independent one conditioned on the noisy observation. This result leads to a full asymptotic characterization of the $k$th-order joint empirical distribution of $(\X^n, \Z^n, \Y^n)$ and, for any fixed k, an exact \blue{asymptotic} expression for the achievable loss, which substantially improves upon the bound established in~\cite{weissman_empirical_2005} for additive noise channels.

\blue{
Our proposed scheme has the following benefits as compared to a direct denoising approach.
The compression yields a representation of $\Z$ at rate around $\lim_n \frac{1}{n}
I(\X^n; \Z^n)$, which is advantageous in rate-limited scenarios where the observations must be stored or communicated.
Denoising approaches, such as the DUDE \cite{weissman_universal_2005} or simply outputting the Bayes response offer no guarantee on the rate required to store or communicate the denoised output.
In principle, our scheme can operate without knowledge of the distribution of the source just like the DUDE, e.g. by using fixed-distortion Yang-Kieffer codes \cite{yang_simple_1996} as the lossy compressor, although in this paper we work with the standard assumption that the distribution of the source $X^n$ and $P_{Z|X}$ are known.
Additionally, one consequence of our framework is that reconstructions $Y^n$ have the same distribution as the source without being separately imposed as a design constraint. This is akin to perfect perception in the recent rate-distortion-perception literature \cite{chen_information_2024, blau_rethinking_2019}.
Finally, using compression as a tool for denoising opens the way to leveraging practical, off-the-shelf lossy compressors, including recent advances in neural compression \cite{chen_information_2024, balle_nonlinear_2021}. Compression-based denoising allows us to use architectures and techniques from neural compressors without modification for the purpose of denoising, with the added benefits of a rate-limited representation and perfect perception.
}

\subsection{Organization}
In \Cref{sec:prelim} we introduce the problem and give some relevant known results.
In \Cref{sec:meat} we first extend results on compression-based denoisers to non-additive noise channels and develop the results that give the exact \blue{asymptotic} characterization of the loss of the compression-based denoiser.
In \Cref{sec:apps} examine a few special cases of our setting to demonstrate the improvement of our characterization and make some comparisons to related settings.
All skipped proofs appear in \Cref{sec:proofs}. 

\section{Preliminaries} \label{sec:prelim}
    \subsection{Notation and Conventions}
We define $[n]:= \set{1, \dots, n}$.
We notate contiguous subsequences $X_m^n:=(X_{m}, X_{m+1}, \dots, X_{n-1}, X_n)$, and denote $X^n := X_1^n$.
We denote the law of some random variable $V$ by $\thedist_{V}$.
For convenience, we sometimes write $U \disteq V$ to mean equality in distribution $\thedist_{V} = \thedist_{U}$.
For measures $P,Q$ such that $P\ll Q$, we denote the Radon-Nikodym derivative by $\frac{dP}{dQ}$.
If $V$ takes values on a finite alphabet, the p.m.f. is denoted by $\thedensity_V$, and similarly $\thedensity_{V|U}$ for conditional p.m.f.s. 

For a probability measure $P$ and Markov kernel $Q$, we denote the induced joint distribution by $Q\otimes P$, and the induced marginal in the first coordinate by $Q\circ P$.
We use an exponent to denote the product measure of a measure with itself, e.g. $P^2 = P\times P$.
For example, we have $P_{X|Y}\otimes P_Y = P_{X,Y}$ and $P_{X|Y}\circ P_Y = P_X$.
For distributions $P,Q$, we denote the total variation distance by $\tv{P-Q}$ and the relative entropy by $\fdiv{P}{Q}$.

We denote the entropy of a random variable $V$ as $H(V)$, and the entropy rate of a random process $\seqfont{V}$ as $\mathbb{H}(V) := \lim_{n\to\infty}\frac{1}{n} H\left(V^n\right)$ when it exists.

\subsection{Problem Setting}\label{sec:setting}
\newcommand{\x}{x}
\newcommand{\xc}{\alphfont{\X}}
\newcommand{\z}{z}
\newcommand{\zc}{\alphfont{\Z}}
\newcommand{\y}{y}
\newcommand{\yc}{\alphfont{\Y}}
\newcommand{\ys}{\seqfont{\Y}}

\newcommand{\pzgx}{\thedensity_{\Z|\X}}
\newcommand{\pzgxk}{\thedensity_{\Z^k|\X^k}}
\newcommand{\pzgy}{\thedensity_{\Z|\Y}}
\newcommand{\pzgyk}{\thedensity_{\Z^k|\Y^k}}
\newcommand{\Pygz}{\thedist_{\Y|\Z}}
\newcommand{\Pzgy}{\thedist_{\Z|\Y}}

Throughout, let $\xs = (\X_1, \X_2, \dots)$ denote a stationary ergodic process taking values in the alphabet $\xc$.
Similarly let $\zs$ be a random process taking values in the alphabet $\zc$.
We assume a known memoryless channel $\Pzgx$, such that $\zs$ is produced by passing $\xs$ through the channel.

For each $n$, let $\Y^n$ be an $n$-tuple of random variables \blue{which may take} values in $\yc^n$.
Unless otherwise specified, $n$ will be the length of the block in consideration and $k\in \nats$ will be such that $1\leq k \leq n$.
Note here there need not be some process $\ys$ which agrees with all $\Y^n$, even in distribution.
Unless otherwise specified, for all $n$ we have the Markov chain
\begin{equation}
    \label{eq:thesetup}
    \X^n\mchain\Z^n\mchain\Y^n.
\end{equation}
The goal is to design a (possibly randomized) mapping $\Z^n\to\Y^n$ depending \emph{only} on $\Pzgx$ and possibly the distribution of $\xs$, such that $\Y^n$ recovers $\X^n$ well.
\blue{We will assume that $\yc \supseteq \xc$, i.e., that denoising outputs are at least permitted to take any values that the source itself can take.}
We will consider a loss function for the recovery $\Lambda: \xc\times\yc \to [0, \Lambda_{\max}]$, and define
\begin{equation}
    \Lambda_n(\X^n, \Y^n) = \frac{1}{n}\sum_{i=1}^n \Lambda(\X_i, \Y_i).
\end{equation}

We will show a sense in which a sequence of good lossy source codes $\set{\Y^n\left(\cdot\right)}_n$ designed for a certain distortion measure $\loss$ that depends only on the channel $P_{Z|X}$ is also good for the denoising task under any reasonable loss function $\Lambda: \xc\times\yc \to [0, \Lambda_{\max}]$.

The notion of a ``good'' lossy source code is formalized  through the following definitions.

\begin{definition}
    For a fixed single-letter distortion measure $\loss: \zc\times\yc\to [0, \infty]$, we define the distortion of a block $(z^n,y^n)$ as
    \begin{equation}
        \label{eq:blocklossdef}
        \loss_n\left(\z^n, \y^n\right) = \frac{1}{n}\sum_{i=1}^n \loss\left(\z_i, \y_i\right).
    \end{equation}
    We denote the rate-distortion function
    \begin{subequations}
    \begin{align}
        \label{eq:rddef}
        R\left(\Z^k, D\right) &= \inf_{\thedist_{\Y^k| \Z^k}:\exof{\loss_k\left(\Z^k, \Y^k\right)} \leq D} \frac{1}{k}I\left(\Z^k; \Y^k\right)\\
        R\left(\zs, D\right) &= \lim_{k\to\infty}R\left(\Z^k, D\right).
    \end{align}
    \end{subequations}
\end{definition}
Hereafter, unless otherwise specified, $\Y^n$ is the reconstruction sequence of a lossy source code for $\Z^n$.

\begin{definition}
    For a fixed $n$, a \emph{code} consists of a codebook $\mathcal{C}\subseteq \yc^n$ and a mapping $\phi: \zc^n\to \mathcal{C}$.
    We define the rate of the code to be
    \begin{equation}
        R = \frac{1}{n}\log\abs{\mathcal{C}}.
    \end{equation}
    \blue{Here, and throughout, all $\log$ and $\exp$ are taken with respect to the same common base.}
    A sequence of codes $(\mathcal{C}_n, \phi_n)$ is called \emph{good} at some $(R, D)$ on the rate-distortion curve if the rate is bounded by $R$, i.e.
    \begin{equation}
        \limsup_{n\to\infty} \frac{1}{n}\log\abs{\mathcal{C}_n} \leq R,
    \end{equation}
    and the distortion satisfies
    \begin{equation}
        \limsup_{n\to\infty} \exof{\loss_n\left(\Z^n, \phi_n\left(\Z^n\right)\right)} \leq D.
    \end{equation}
When a sequence of codes $(\mathcal{C}_n, \phi_n)$ is good at $(R,D)$, we will, refer to the corresponding decoder output $\Y^n = \phi_n(\Z^n)$ as good for the same $(R,D)$.
\end{definition}
The notion of goodness defined above can equivalently be expressed as a condition solely on the reconstructions $\Y^n = \phi_n(\Z^n)$, as follows.
\begin{definition}
    \label{def:good}
    A sequence of reconstructions $\set{\Y^n\left(\cdot\right)}_n$ is called \emph{good} at some $(R, D)$ on the rate-distortion function $R\left(\zs, D\right)$ if $\frac{1}{n}H\left(\Y^n\right) \leq R$ and
    \begin{equation}
        \limsup_{n\to\infty} \exof{\loss_n\left(\Z^n, \Y^n\left(\Z^n\right)\right)} \leq D.
    \end{equation}
\end{definition}
Henceforth we will talk about good sequences $\Y^n$ without explicitly referring to the underlying sequence of lossy source codes.

The following are needed to state our results.
\newcommand{\optformat}[1]{\tilde{#1}}
\newcommand{\zopt}{\optformat{\Z}}
\newcommand{\yopt}{\optformat{\Y}}
\newcommand{\znk}{\tilde{\Z}^{(n,k)}}
\newcommand{\ynk}{\tilde{\Y}^{(n,k)}}
\begin{definition}
    \blue{For $n > 2k$}, we denote the \emph{empirical distribution}, which is a function of $(\X^n, \Z^n, \Y^n)$, by
    \begin{equation}
        \begin{split}
        &\empirical{\X^n, \Z^n, \Y^n}\left(\x_0, \z_{-k}^k, \y_{-k}^k\right)\\
        &= \frac{1}{n-2k}\sum_{i=k+1}^{n-k} \ind{\X_i = x_0, \Z_{i-k}^{i+k} =\z_{-k}^k, \Y_{i-k}^{i+k} =\y_{-k}^k},
        \end{split}
    \end{equation}
    where $\ind{\cdot}$ is the indicator function.
    For finite alphabets, $\empirical{\X^n, \Z^n, \Y^n}$ can be identified with a random vector in $\reals^{\abs{\xc}\abs{\yc}^{2k+1}\abs{\zc}^{2k+1}}$ that always lies on the probability simplex.
    Expectations can be taken the standard way for random vectors, and will always result in a valid distribution.
    We define $Q\idx{n}$ to be a distribution on $\xc, \zc^{2k+1}, \yc^{2k+1}$ by taking the expectation of $\empirical{\X^n, \Z^n, \Y^n}$, and use subscripts to denote the corresponding marginal or conditional p.m.f.s obtained from the joint distribution $Q\idx{n}$.
    More explicitly,
    \begin{subequations}
    \begin{align}
        Q\idx{n}\left(\x_0, \z_{-k}^k, \y_{-k}^k\right)
        &= \exof{Q\left[\X^n, \Z^n, \Y^n\right]\left(\x_0, \z_{-k}^k, \y_{-k}^k\right)}\\
        Q\idx{n}_{Z_{-k}^k, Y_{-k}^k}\left(\z_{-k}^k, \y_{-k}^k\right) &= \sum_{\x_0} \exof{Q\left[\X^n, \Z^n, \Y^n\right]\left(\x_0, \z_{-k}^k, \y_{-k}^k\right)}\\
        Q\idx{n}_{X_0|Z_{-k}^k, Y_{-k}^k}\left(\x_0, \z_{-k}^k, \y_{-k}^k\right) &= 
        \frac{\exof{Q\left[\X^n, \Z^n, \Y^n\right]\left(\x_0, \z_{-k}^k, \y_{-k}^k\right)}}{ \sum_{\x_0} \exof{Q\left[\X^n, \Z^n, \Y^n\right]\left(\x_0, \z_{-k}^k, \y_{-k}^k\right)}}.
    \end{align}
    \end{subequations}
    We note that $Q\idx{n}$ is a valid distribution by linearity of expectation, and conditioning and marginalization of $Q\idx{n}$ occur after taking expectations.
    Convergence of a sequence of $Q\idx{n}$ simply means convergence of a sequence of real vectors in $\reals^d$ for appropriate $d$, or, equivalently, convergence in distribution of random variables drawn according to the distribution.
\end{definition}

\subsection{Prior Results}

If $\set{\Y^n\left(\cdot\right)}_n$ induces a $\thedist_{\Y^n| \Z^n}$  that achieves $R\left(\Z^n, D_n\right) = R^*$ for some fixed $R^*$, then $\set{\Y^n\left(\cdot\right)}_n$ is good in the sense of \Cref{def:good}.
Often a sort of converse result is true: any sequence of good codes must have empirical distribution approaching \emph{the} distribution that achieves the rate-distortion function.
Sufficient conditions are given in the following result.
\begin{theorem}[Theorem 3 of \cite{weissman_empirical_2005}]
\label{thm:three}
    \assumefinitealphabet{}
    Suppose the sequence of \blue{reconstructions} $\set{\Y^n\left(\cdot\right)}_n$ is good at $(R\left(\zs, D\right), D)$. 
    Suppose the condition
    \begin{equation}
    \label{eq:threecond}
        R\left(\Z^k, D\right) = R\left(\zs, D\right) + \frac{1}{k}H\left(\Z^k\right) - \mathbb{H}\left(\zs\right)
    \end{equation}
    holds, and suppose that $R\left(\Z^k, D\right)$ is uniquely achieved by the distribution of the pair $\left(\zopt^k, \yopt^k\right)$.
    Then,
    \begin{equation}
        Q\idx{n}_{\Z^k, \Y^k}
        \to \thedist_{\zopt^k, \yopt^k}\text{ as } n\to\infty.
    \end{equation}
\end{theorem}
When the $\X_i$ (and therefore $\Z_i$) are i.i.d., assumption \cref{eq:threecond} clearly holds.
The following result shows that \cref{eq:threecond} can be satisfied if the distortion measure and level are matched to the noise channel in the special case of additive noise channels.
\begin{theorem}[Theorem 4 of \cite{weissman_empirical_2005}]
\label{thm:four}
    Suppose $\xc=\zc=\yc$ is a finite abelian group, with group operation denoted $+$.
    Suppose $\zs$ is the result of additive white noise applied to $\xs$, i.e.
    \begin{equation}
        \Z_i = \X_i + N_i
    \end{equation}
    for i.i.d. $N_i$.
    If we choose the difference distortion measure $\loss(\z,\y) = -\log\thedensity_{N}\left(\z-\y\right)$,
    the rate-distortion function has the form given by
    \begin{equation}
        \label{eq:oldrdexpression}
        R\left(\Z^k, H(N)\right) = \frac{1}{k}H\left(\Z^k\right) - H\left(N\right),
    \end{equation}
    which is achieved by $\left(\Z^k, \Y^k\right) \disteq \left(\Z^k, \X^k\right)$, uniquely when the channel matrix of $\Pzgx$ is invertible.
\end{theorem}

\bigbreak
We see that by taking the limit on both sides of \cref{eq:oldrdexpression}, we have
\begin{equation}
    R\left(\zs, D\right) = \lim_{k \to \infty}R\left(\Z^k, H\left(N\right)\right) = \lim_{k \to \infty} \frac{1}{k}H\left(\Z^k\right) - H\left(N\right) = \mathbb{H}\left(\zs\right) - H\left(N\right).
\end{equation}
Subtracting from \cref{eq:oldrdexpression} yields
\begin{equation}
    R\left(\Z^k, H\left(N\right)\right) - R\left(\zs, D\right) = \frac{1}{k}H\left(\Z^k\right) - H\left(\zs\right),
\end{equation}
which is exactly the condition \cref{eq:threecond}.
Applying \Cref{thm:three}, we arrive at the following corollary. 
\begin{cor}
\label{cor:old}
    Under the assumptions of \Cref{thm:four}, if the channel matrix of $\Pzgx$ is invertible, and $\set{\Y^n}_n$ is a sequence of good codes at distortion level $H(N)$, \blue{then we obtain}
    \begin{equation}
    \label{eq:denoiseroutputdist}
        Q\idx{n}_{\Z^k, \Y^k}
        \to \thedist_{\Z^k, \X^k}\text{ as } n\to\infty.
    \end{equation}
\end{cor}

\bigbreak
We can then use good codes $\Y^n$ to estimate the original signal $\X^n$.
In \cite{weissman_empirical_2005}, the following bound on denoising performance is derived.
\begin{theorem}[Theorem 5 of \cite{weissman_empirical_2005}]
\label{thm:five}
    Under the conditions of \Cref{thm:four}, if $\set{\Y^n}_n$ is a sequence of good codes for $\zs$ at distortion level $H(N)$, then for any loss function $\Lambda: \xc\times\yc \to [0, \Lambda_{\max}]$ the following holds
    \begin{equation}
    \label{eq:oldupperbound}
        \begin{split}&\limsup_{n\to\infty}\exof{\Lambda_n\left(X^n, Y^n(Z^n)\right)} \\
        &\leq \exof[Z_{-\infty}^\infty]{\sup\set{\exof{\Lambda(U,V)}:U\sim\thedist_{X_0|\Z_{-\infty}^\infty}, V\sim\thedist_{X_0|\Z_{-\infty}^\infty}}}.
    \end{split}
    \end{equation}
\end{theorem}

\newcommand{\ird}{R_{I}}
\newcommand{\distdist}{d}
{
\subsection{Indirect Rate-Distortion}
We now define the indirect rate-distortion problem. This is a classical variation of the standard rate-distortion problem where the source $\xs$ is observed through a noisy channel.
The problem has been introduced in \cite{dobrushin_information_1962} and further developed in \cite{witsenhausen_indirect_1980}.
We again contrast our main problem setting with indirect rate-distortion by noting that in indirect rate-distortion, the compressor is optimized for known distortion measure defined on $\xc\times \yc$, while in our setting no such distortion measure is assumed.

We define the \emph{indirect rate-distortion} curve, denoted by $\ird(L)$, for the distortion $\Lambda$ as the set of optimal values of the following optimization problem, parameterized by the loss $L$.
\begin{align}
\min_{\thedist_{\Y^k|\Z^k}} &\blue{\frac{1}{k}}I(\Z^k;\Y^k)\\
\text{subject to} \quad &\exof{\Lambda_k(\blue{\X^k, \Y^k})} \leq L.
\end{align}
Here we maintain the assumptions that $\X^k\mchain\Z^k\mchain\Y^k$ form a Markov chain and still have a fixed and known channel $\Pzgx$ as in the original setup.
The only difference in this setting is that $\Lambda$ is known ahead of time and the compressor is designed to minimize this particular distortion measure.

As shown by Witsenhausen in \cite{witsenhausen_indirect_1980}, the indirect rate-distortion can be reduced to rate-distortion for distortion measure given by
\begin{equation}
\distdist(\z,\y) = \exof{\Lambda(\X,\y)\given Z=z}.
\end{equation}
} 

\section{Main Results} \label{sec:meat}
    We generalize and strengthen the results of \cite{weissman_empirical_2005} in two essential ways.
First, we show that we can use good lossy source codes to do denoising more generally; the DMC does not need to be an additive noise channel.
Second, we give an \emph{exact} characterization of the denoising performance of said compressors in lieu of the upper bound given in \cite{weissman_empirical_2005}, which, we will demonstrate in \ref{sec:apps}, can be quite loose.
\subsection{Denoising for General Noise Channels}
It is natural to consider when the condition \cref{eq:threecond} for \Cref{thm:three} might hold if $\loss$ is not a difference distortion measure.
To this end, we choose our distortion measure
\begin{equation}
    \label{eq:lossdef}
    \loss(\z,\y) = -\log\pzgx\left(\z\given \y\right),
\end{equation}
\blue{and choose compressors that are restricted to output symbols in $\xc$. Thus we assume $\yc = \xc$ for the remainder of the section, which is a more restrictive setting than allowing a larger $\yc$.}
For a given observation $\z$, the distortion is minimized when the reconstruction $\y$ is the value of $\X$ that best explains the observation, in the sense of having maximum likelihood.
In the case of additive noise $\Z_i = \X_i + N_i$, our choice recovers the distortion $\loss(\z,\y) = -\log\thedensity_{N}(\z-\y)$ from \cite{weissman_empirical_2005}.
Since we aim to recover $\xs$, one may guess that it is appropriate to target the distortion that would be achieved if $\Y$ \blue{were} distributed as $\X$. Thus, we choose
\begin{equation}
    D = H\left(\Z\given \X\right).
\end{equation}
With the above choice of distortion, \Cref{thm:four} generalizes naturally as follows.
\begin{theorem}
    \label{thm:thmfour}
    Suppose the alphabets $\xc, \zc, \yc$ are finite.
    Under the distortion \eqref{eq:lossdef}, the rate-distortion function has the form given by
    \begin{equation}
    \label{eq:newrdexpression}
        R\left(\Z^k, H\left(\Z\given \X\right)\right) = \frac{1}{k}I\left(\Z^k; \X^k\right) = \frac{1}{k}H\left(\Z^k\right) - H\left(\Z\given \X\right).
    \end{equation}
    The rate-distortion function is achieved when $\left(\Z^k, \Y^k\right) \disteq \left(\Z^k, \X^k\right)$, uniquely so if the channel matrix of $\Pzgx$ is of full row rank.
\end{theorem}
\begin{IEEEproof}[Proof of \Cref{thm:thmfour}]
    Let $\Pygz$ be feasible, i.e.
    \begin{equation}
        \label{eq:rdoptconstraint}
        \exof{\loss_k\left(\Z^k, \Y^k\right)} \leq H\left(\Z\given \X\right).
    \end{equation}
    Using the fact \blue{that} the channel is memoryless,
    \begin{align}
        \exof{-\log\pzgxk\left(\Z^k\given \Y^k\right)} \leq H\left(\Z^k\given \X^k\right)
    \end{align}

    Now,
    \begin{align}
        &I\left(\Z^k; \Y^k\right)\\
        &= H\left(\Z^k\right) - H\left(\Z^k\given \Y^k\right)\\
        &= H\left(\Z^k\right) - \exof{-\log\pzgyk\left(\Z^k\given \Y^k\right)}\\
        \begin{split}
        &= H\left(\Z^k\right) - \exof{-\log\pzgxk\left(\Z^k\given \Y^k\right)}\\
        &+ \left(\exof{\log\pzgyk\left(Z^k\given Y^k\right)} - \exof{\log\pzgxk\left(\Z^k\given \Y^k\right)}\right)
        \end{split}\\
        \begin{split}
            &= H\left(\Z^k\right) - \exof{-\log\pzgxk\left(\Z^k\given \Y^k\right)} + \\
            &\exof{\fdiv{\thedist_{\Z^k | \Y^k}\left(\cdot\given \Y^k\right)}{\thedist_{\Z^k | \X^k}\left(\cdot\given \Y^k\right)}}
        \end{split}\\
        &\geq H\left(\Z^k\right) - H\left(\Z^k\given \X^k\right).
    \end{align}

    Substituting $\Y = \X$, we see that when $\left(\Z^k, \Y^k\right) \disteq \left(\Z^k, \X^k\right)$, the inequality is met with equality and the constraint \cref{eq:rdoptconstraint} is met with equality.

    Since rate-distortion functions are achieved at a unique backward channel $\thedist_{\Z^k|\Y^k}^*$ \blue{for finite alphabets} \cite[Section 1.3, Problem 3]{csiszar_information_1981}, see also \cite[Theorem 9.4.1]{gallager_information_1968},
    the joint distribution is unique by the rank assumption on the channel matrix of $\Pzgx$.
\end{IEEEproof}

Similar to the derivation of \Cref{cor:old}, we can take the limit on both sides of \cref{eq:newrdexpression} for
\begin{equation}
    R\left(\zs, D\right) = \lim_{k \to \infty}R\left(\Z^k, H\left(\Z\given \X\right)\right) = \lim_{k \to \infty} \frac{1}{k}H\left(\Z^k\right) - H\left(\Z\given \X\right) = \mathbb{H}\left(\zs\right) - H\left(\Z\given \X\right).
\end{equation}
Subtracting from \cref{eq:newrdexpression} yields
\begin{equation}
    R\left(\Z^k, H\left(\Z\given \X\right)\right) - R\left(\zs, D\right) = \frac{1}{k}H\left(\Z^k\right) - \mathbb{H}\left(\zs\right),
\end{equation}
which is again the condition \cref{eq:threecond}.
Thus, we can apply \blue{\Cref{thm:three}}.
\begin{cor}
\label{cor:general}
    Under the assumptions of \Cref{thm:thmfour}, if the channel matrix of $\Pzgx$ has full row rank, and $\set{\Y^n}_n$ is a sequence of good codes,
    \begin{equation}
        Q\idx{n}_{\Z^k, \Y^k}
        \to \thedist_{\Z^k, \X^k}\text{ as } n\to\infty.
    \end{equation}
\end{cor}
Thus, even when $\Pzgx$ is not an additive noise channel, our choice of the distortion measure and the distortion level guarantees that lossy compression of the observation asymptotically samples from the posterior distribution $\thedist_{\X^k|\zs}$ of the signal.
In applications, a way to sample from the posterior can itself be of interest.
In the next section,
we further specify the behavior of the reconstructions, and we will characterize its denoising performance with respect to a loss $\Lambda$.

\subsection{Denoising Performance}
\newcommand{\mixcoeff}{\delta}
From the previous section we have a characterization of the asymptotic behavior of $Q\idx{n}_{\Z^k, \Y^k}$.
The denoising performance depends on the joint distribution of source, observation, and reconstruction $\xs, \zs, \ys$, which motivates the following results.
Recall that we have the Markov chain $\X^n\mchain\Z^n\mchain\Y^n$.
However, for fixed $n,k$, the Markov relation $\X^k\mchain\Z^k\mchain\Y^k$ for $k<n$ does not hold in general, due to the ``memory'' in $\xs$.
We next show that the \emph{empirical} joint distribution of the source,
the observation, and the denoiser output asymptotically satisfies this Markov condition.
We require the following mild assumption on $\xs, \zs$.
\blue{Henceforth we will consider double-sided extensions of the processes $(\xs, \zs)$, which will exist and still be ergodic.}
\begin{definition}
    Suppose $\xs, \zs$ are jointly stationary.
    We define their \emph{double-sided mixing coefficient} as
    \begin{equation}
    \begin{split}
        \mixcoeff_k\left(\xs, \zs\right)=\\
        \esssup_{\Z_{-\infty}^{k-1}, \Z_{k+1}^{\infty}} \max_{\x_0, \z_{-k}^k} \big|&\thedist_{\X_0|\Z_{-k}^k}\left(\x_0\given \z_{-k}^k\right) \\
        &- \thedist_{\X_0|\zs}\left(\x_0\given \z_{-k}^k,\Z_{-\infty}^{k-1}, \Z_{k+1}^{\infty}\right)\big| 
    \end{split}
    \end{equation}
    and say that $(\xs, \zs)$ are \emph{double-sided mixing} if additionally
    \begin{equation}
        \lim_{k\to \infty}\mixcoeff_k\left(\xs, \zs\right) = 0.
    \end{equation}
    Hereafter we use $\mixcoeff_k := \mixcoeff_k\left(\xs, \zs\right)$ for brevity.
\end{definition}

Despite the fact that $\thedist_{\X_0|\Z_{-k}^k}\left(x_0\given \Z_{-k}^k\right) \toae \thedist_{\X_0|\Z_{-k}^k}\left(x_0\given \Z_{-\infty}^\infty \right)$ (by the martingale convergence theorem), it is not hard to construct processes for which the $\mixcoeff_k$ never vanish.
However, the class of pairs of processes that are double-sided mixing is large and arguably includes all those of practical interest.
For example, for reasonable $\Pzgx$, requiring $\xs$ to be a Markov chain is more than enough:
\begin{prop}
    \label{prop:markovmixing}
    \assumefinitealphabet{}
    Suppose $\xs$ is an ergodic Markov chain, and suppose $\Pzgx\left(\z\given \x\right) > 0$ for all $\z, \x$.
    Then, $\mixcoeff_k\to 0$ exponentially fast.
\end{prop}
This result can be extended to the case when $\xs$ is an order-$m$ Markov process.

Roughly speaking, the following claim establishes that the double-sided mixing coefficient controls the extent to which the empirical distribution violates the Markov condition $\X_0\mchain\Z_{-k}^k \mchain \Y_{-k}^k$.
\begin{lemma}
\label{lemma:mixingupperbound}
    Suppose $\xs, \zs$ are jointly stationary.
    Let $n,k\in \nats$.
    Then
    \begin{equation}
        \tv{Q\idx{n}_{X_0|Z_{-k}^k, Y_{-k}^k} - \thedist_{X_0|Z_{-k}^k}} \leq \abs{\xc} \mixcoeff_k.
    \end{equation}
\end{lemma}

We know from \Cref{cor:general} that, conditioned on the observations, a source symbol $\X_i$ and its reconstruction $\Y_i$ are both distributed according to the posterior.
Applying \Cref{lemma:mixingupperbound} allows us to additionally deduce they are essentially conditionally independent which, in turn, leads to the complete characterization of the denoising performance in the following theorem.
\begin{theorem}
\label{thm:denoisermixingcoeff}
    \assumefinitealphabet{}
    Suppose $(\xs, \zs)$ are double-sided mixing, and suppose the channel matrix of $\Pzgx$ \blue{has full row rank}.
    Let $\Lambda: \xc\times\yc \to [0, \Lambda_{\max}]$ be a loss function as defined in \Cref{sec:setting}.
    Suppose $\set{\Y^n}_n$ is a sequence of good codes for $\zs$ under distortion $\loss(\z,\y) = -\log\pzgx\left(\z\given \y\right)$ at distortion level $H\left(\Z\given \X\right)$.
    Then,
    \begin{equation}
    \begin{split}
        &\lim_{n\to\infty}\exof{\Lambda_n\left(X^n, Y^n(Z^n)\right)} \\
        &= \exof[\zs]{\exof[(U,V)\sim\left(\thedist_{X_0|\zs}\right)^2]{\Lambda(U, V)}}.
    \end{split}
    \end{equation}
\end{theorem}
This theorem constitutes a complete characterization of the loss achieved.
The value on the right hand side improves upon the upper bound in \Cref{thm:five} of \cite{weissman_empirical_2005} in that instead of using the worst case coupling, $U,V$ are assumed independent. 
\begin{IEEEproof}[Proof of \Cref{thm:denoisermixingcoeff}]
    The conditions for \Cref{thm:thmfour,lemma:mixingupperbound} are satisfied.
    We use the fact that the expectation of the loss is determined by the empirical distribution.
    \begin{align}
        &\exof{\Lambda_n\left(X^n, Y^n(Z^n\right)}\\
        &= \exof{\frac{1}{n}\sum_{x, y}\Lambda(x,y)\sum_i\ind{\X_i = x, \Y_i = y}}\\
        &= \exof[(X,Y)\sim Q\idx{n}_{X_0, Y_0}]{\Lambda(X, Y)}\\
        \begin{split}
        &= \exof[\tilde{Z}_{-k}^k,\tilde{Y}_{-k}^k\sim Q\idx{n}_{{Z}_{-k}^k, {Y}_{-k}^k}]{\exof[\tilde{X}\sim Q\idx{n}_{X_0|Z_{-k}^k, Y_{-k}^k}]{\Lambda(\tilde{X}, \tilde{Y}_0)}}\\
        \end{split}\\
        \label{eq:lossproof2}
        \begin{split}
        &= \exof[\tilde{Z}_{-k}^k,\tilde{Y}_{-k}^k\sim Q\idx{n}_{{Z}_{-k}^k, {Y}_{-k}^k}]{\exof[\tilde{X}\sim \thedist_{X_0|Z_{-k}^k}]{\Lambda(\tilde{X}, \tilde{Y}_0)}}\\
        &\qquad +  \Lambda_{\max}\abs{\xc}\mixcoeff_k
        \end{split}\\
        \label{eq:lossproof3}
        \begin{split}
        &=  \exof[U,V\sim\thedist_{X_0|\Z_{-k}^k}\text{ i.i.d.}]{\Lambda(U, V)}\\
        &\qquad + o_n(1) +  \Lambda_{\max}\abs{\xc}\mixcoeff_k + o(1/n)
        \end{split}\\
        &= \exof[U,V\sim\thedist_{X_0|\Z_{-k}^k}\text{ i.i.d.}]{\Lambda(U, V)} + o_n(1) + o_k(1),
    \end{align}
    where
    \eqref{eq:lossproof2} follows from \Cref{lemma:mixingupperbound}, and \eqref{eq:lossproof3} follows from \Cref{cor:general}.
    \blue{Here, $o_n(f(n))$ denotes a quantity $g(n)$ such that $\lim_{n\to\infty} g(n)/f(n) = 0$, and $o_k(f(k))$ is defined analogously for $k$.}
    Taking the limit in $n$ and then $k$ finishes the proof.
\end{IEEEproof}

\blue{
\begin{cor}\label{cor:metric}
    Let $\Lambda$ be a metric and let $\hat{X}(\zs) = \operatornamewithlimits{argmin}_x \exof{\Lambda(X_0, x)\given \zs} $ be the Bayes response.
    \begin{equation}
        \lim_{n\to\infty}\exof{\Lambda_n\left(X^n, Y^n(Z^n)\right)}
        \leq 2\exof{\Lambda(X_0, \hat X(\zs))},
    \end{equation}
\end{cor}
\begin{IEEEproof}
    From the triangle inequality we have
    \begin{align}
        &\lim_{n\to\infty}\exof{\Lambda_n\left(X^n, Y^n(Z^n)\right)} \\
        &= \exof[\zs]{\exof[(U,V)\sim\left(\thedist_{X_0|\zs}\right)^2]{\Lambda(U, V)}}\\
        &\leq \exof[\zs]{\exof[(U,V)\sim\left(\thedist_{X_0|\zs}\right)^2]{\Lambda(U, \hat X(\zs)) + \Lambda(\hat X(\zs), V)}}\\
        &= 2\exof{\Lambda(X_0, \hat X(\zs))},
    \end{align}
\end{IEEEproof}
Thus, when the loss is a metric, our scheme never incurs more than twice the Bayes optimal loss.
We note that $\hat X$ has no constraint on rate.
}

\section{Special Cases} \label{sec:apps}
    We next show that when the loss is mean squared error (MSE), \Cref{thm:denoisermixingcoeff} gives a factor of 2 improvement over the bound in \cite{weissman_empirical_2005} (reproduced as \Cref{thm:five} here).
Note that \Cref{thm:denoisermixingcoeff} also applies for channels that are not additive noise.
\begin{example}[MSE]\label{ex:mse}
    \blue{Let $\xc$ be a finite cardinality subset of $\yc = \reals$ and $\Lambda(\x,\y) = \left(\x-\y\right)^2$.}
    Let $\xs$ be an ergodic process and let $\Pzgx$ be a DMC with invertible channel matrix such that $(\xs, \zs)$ is double-sided mixing.
    Let $\set{\Y^n\left(\cdot\right)}_n$ be a sequence of good codes for $\zs$ for distortion measure $\loss(\z,\y) = -\log\pzgx\left(\z\given \y\right)$ at distortion level $D = H\left(Z\given X\right)$.

    The Bayes optimal denoiser $\hat{X}$ outputs
    \begin{equation}
        \hat{X}_i(\zs) = \exof{X_i\given \zs},
    \end{equation}
    which achieves the MSE
    \begin{equation}
        \lim_{n\to\infty}\exof{\Lambda_n\left(X^n, \left(\hat{X}(\zs)\right)_{1}^n\right)} = \exof{\varof{X_0\given \zs}}.
    \end{equation}
    Applying \Cref{thm:denoisermixingcoeff}, we conclude that using the compressor-based denoising achieves MSE
    \begin{align}
        &\lim_{n\to\infty}\exof{\Lambda_n\left(X^n, Y^n(Z^n)\right)}\\
        &= \exof{2\varof{X_0\given \zs}}.
    \end{align}
    On the other hand, \Cref{thm:five} (from \cite{weissman_empirical_2005}) gives the upper bound
    \begin{align}
        &\lim_{n\to\infty}\exof{\Lambda_n\left(X^n, Y^n(Z^n)\right)}\\
        &\leq \exof[\zs]{\sup\set{\exof{(U-V)^2}:U, V\sim\thedist_{X_0|\zs}}}\\
        &\leq \exof{4\varof{X_0\given \zs}},
    \end{align}
    where the second inequality is tight if $\thedist_{X_0|\zs}$ is symmetric (as the coupling with $U=-V$ achieves the supremum).
\end{example}

\begin{figure}
    \centering
    \includegraphics[width=0.8\linewidth]{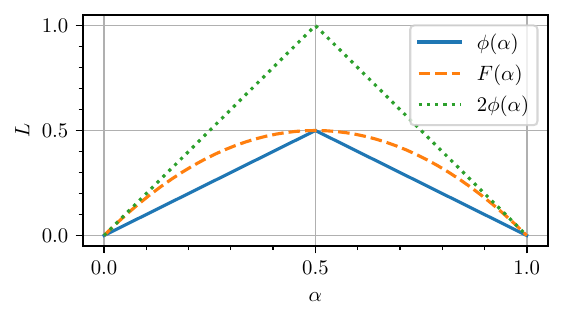}
    \caption{Comparison of the Bayes envelope, compression based denoiser loss, and suboptimal upper bound for denoising a binary $\xs$ passed through a BSC channel under Hamming loss (\Cref{ex:hamming}).}
    \label{fig:bschamming}
\end{figure}

We also apply \Cref{thm:denoisermixingcoeff} to the case of binary sources with Hamming loss. This setup was studied in \cite{donoho_kolmogorov_2002}. In this special case, our denoiser design recovers that of \cite{weissman_empirical_2005}, but we again obtain an improved analysis over \Cref{thm:five}.
\begin{example}[Hamming Distance]
\label{ex:hamming}
    Let $\xc =\zc = \yc = \set{0,1}$.
    Let $\Lambda$ be the Hamming distance.
    Suppose $\xs$ is mixing (and therefore stationary ergodic).
    Let $\zs$ be the result of passing $\xs$ through the DMC given by $\Pzgx = \mathrm{BSC}(D)$.
    These assumptions suffice for $(\xs, \zs)$ to be double-sided mixing.

    Suppose $\set{\Y^n\left(\cdot\right)}_n$ is a sequence of good codes for $\zs$ at Hamming distortion level $D$.
    Applying \Cref{thm:denoisermixingcoeff}, we have
    \begin{equation}
        \lim_{n\to\infty} \exof{\Lambda_n\left(\X^n, \Y^n\right)} = \exof[\zs]{F\left(\pof{X_0=1\given \zs}\right)},
    \end{equation}
    where
    \begin{equation}
        F(\alpha) = 2\alpha(1-\alpha).
    \end{equation}
    For reference, the Bayes optimal denoiser achieves
    \begin{align}
        &\exof[\zs]{\phi\left(\pof{X_0=1\given \zs}\right)}\\
        \text{where}\qquad \phi(\alpha) &= \min(\alpha, 1-\alpha).
    \end{align}
    As in \cite{weissman_empirical_2005}, \Cref{thm:five} yields
    \begin{align}
        &\lim_{n\to\infty}\exof{\Lambda_n\left(X^n, Y^n(Z^n)\right)}\\
        &\leq \exof[\zs]{\sup\set{\exof{\Lambda(U,V)}:U\sim\thedist_{X_0|\zs}, V\sim\thedist_{X_0|\zs}}}\\
        &= \exof[\zs]{2\phi\left(\pof{X_0=1\given \zs}\right)}.
    \end{align}
    where the supremum is achieved by setting
    \begin{align}
        \pof{U=1, V=0} = \pof{U=0, V=1} = \phi(\pof{X_0=1\given \zs})
    \end{align}
    and putting all the remaining probability on $U=V=\operatorname{\arg\max}_x \pof{X_0 = x\given \zs}$.

    We note that
    \begin{equation}
        \label{eq:hammingcompared}
        \phi(\alpha) \leq F(\alpha) \leq 2\phi(\alpha)
    \end{equation}
    for all $\alpha$, and $F(\alpha)= \phi(\alpha)$ whenever $\alpha\in \set{0, \frac{1}{2}, 1}$.
    See \Cref{fig:bschamming} for a comparison of the functions in \cref{eq:hammingcompared}.    
\end{example}
\blue{
\begin{remark}\label{rem:reviewer}
    Considering $\Lambda(x,y) = (x-y)^M$ instead of the MSE in \cref{ex:mse} demonstrates that $\lim_{n\to\infty}\exof{\Lambda_n\left(X^n, Y^n(Z^n)\right)}$ can be as much as $2^M$ times the Bayes envelope.
    For example, this is achieved when $\thedist_{X_0|\zs}$ is uniform over $\set{-1/2, 1/2}$.
    The scheme can still be arbitrarily bad under this denoising loss compared to the Bayes optimum even if $\yc = \xc$, e.g. if $\thedist_{X_0|\zs}$ is uniform over $\set{-1/2, 0, 1/2}$.
    On the other hand, the factor of 2 as seen in \cref{eq:hammingcompared} is the worst possible if $\Lambda$ since the Hamming distance is a metric (\cref{cor:metric}).
\end{remark}
}

To demonstrate the generality of our result, we apply \Cref{thm:denoisermixingcoeff} to a source with memory and a channel that is not an additive noise channel.
\newcommand{\transmat}{M}
\newcommand{\ps}{p_s}
\newcommand{\pe}{p_e}
\newcommand{\tn}{T_{-}}
\newcommand{\tp}{T_{+}}
\newcommand{\atn}{_{-}}
\newcommand{\atp}{_{+}}
\newcommand{\tc}{T_c}
\newcommand{\tf}{T_f}
\newcommand{\atc}{_{c}}
\newcommand{\atf}{_{f}}
\newcommand{\erased}{\epsilon}
\newcommand{\negfactor}[2]{\left(-1\right)^{#1+#2}}
\newcommand{\qlong}{(1-2\ps)}
\newcommand{\q}{q}
\begin{example}[Binary Symmetric Markov Source with Erasures]
\label{ex:bsserasure}
    Let $\xc = \yc = \set{0,1}$ while $\zc = \set{0,1, e}$.
    Let $\Lambda$ be the Hamming distance.
    Let $\xs$ be a binary symmetric Markov source with switching probability $\ps \in(0, \frac{1}{2})$, i.e. the Markov chain with transition probability
    \begin{equation}
        \transmat = 
        \begin{bmatrix}
            1-\ps&\ps\\\ps&1-\ps
        \end{bmatrix}.
    \end{equation}
    Let $\thedist_{\X_0}$ be uniform.
    Let $\Pzgx$ be the erasure channel with erasure probability $\pe\in[0,1)$.
    It can be readily verified that $\xs$ is ergodic and $(\xs, \zs)$ are double-sided mixing.
    It can be seen, e.g. by taking the power of $\transmat$, that for $t\geq s$ we have
    \begin{equation}
    \label{eq:binsourcetimeskip}
        \thedist_{\X_t | \X_s}\left(\x_t\given \x_s\right)
        = \frac{1}{2}\left(\negfactor{\x_t}{\x_s} \qlong^{t-s} + 1\right).
    \end{equation}
    We denote for brevity
    \begin{equation}
        \q = \qlong.
    \end{equation}
    The Bayes-optimal loss is
    \begin{align}
        \frac{\pe \ps}{1-\pe^2\qlong}.
    \end{align}
    The denoising loss can be given by the infinite sum:
    \begin{align}
        &\exof{F\left(\thedist_{\X_0|\zs}\right)}\\
        &= \frac{1}{2}(1-\pe)^2\sum_{s,t\geq 0} \pe^s\pe^{t} \frac{\left(1-\q^{2(t+1)}\right) \left(1-\q^{2s}\right)}{1-\left(\q^{2(t+1)}\right)\left(\q^{2s}\right)}.
    \end{align}
    As the terms in the summation are $O(\pe^{s+t})$, truncation is sufficient for numerical evaluation.
    Details on deriving the above can be found in \Cref{sec:proofs}.
    We see in \Cref{fig:vspe,fig:vsps} that for various parameter values the achieved denoising loss is generally close to the Bayes envelope and that there is a significant improvement over the upper bound from Theorem~\ref{thm:five} due to \cite{weissman_empirical_2005}. 
\end{example}

\begin{figure}
    \centering
    \includegraphics[width=1\linewidth]{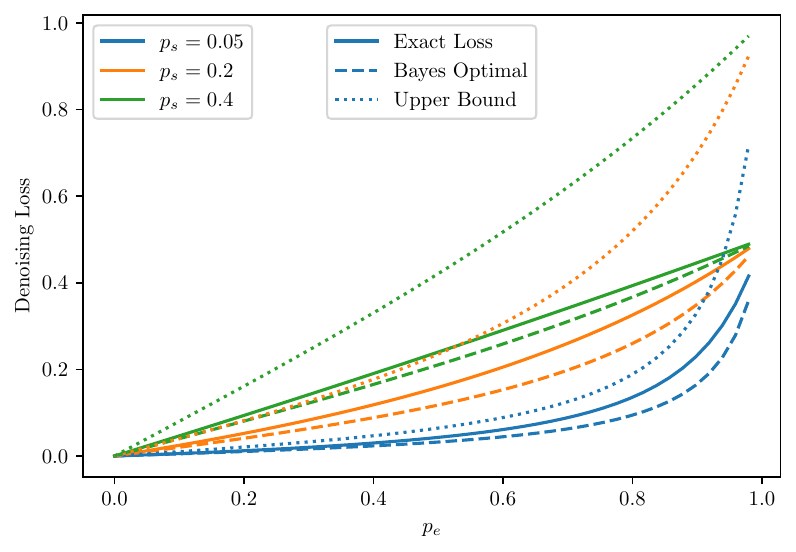}
    \caption{Performance of compression based denoiser in \Cref{ex:bsserasure} for various switching probabilities $\ps$ as a function of the erasure probability $\pe$, compared to the Bayes response and the upper bound \Cref{eq:oldupperbound}. We note that despite the fact that \Cref{thm:five} is necessary to apply the analysis, \Cref{eq:oldupperbound} is still a valid upper bound.}
    \label{fig:vspe}
\end{figure}
\begin{figure}
    \centering
    \includegraphics[width=1\linewidth]{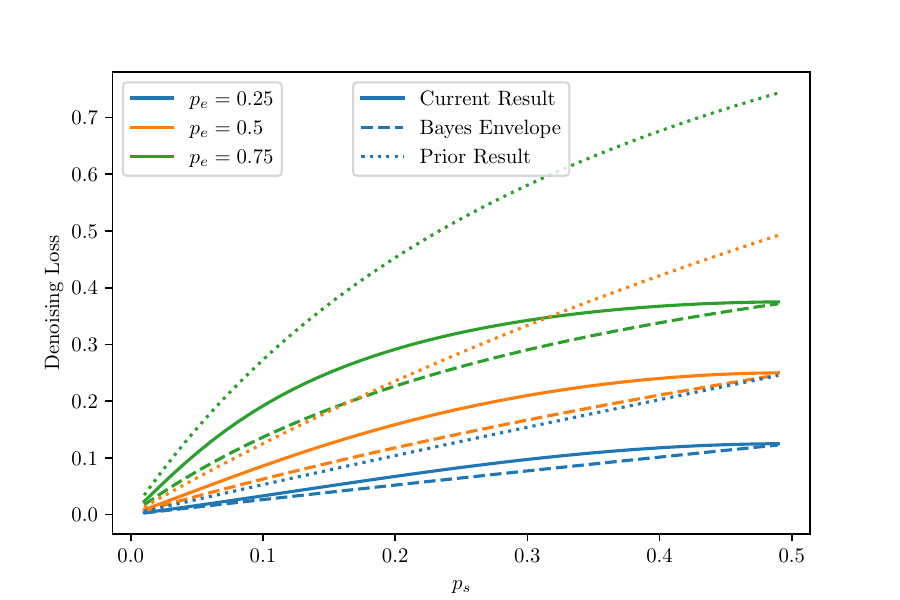}
    \caption{Performance of compression based denoiser in \Cref{ex:bsserasure} for various erasure probabilities $\pe$ as a function of the switching probability $\ps$, compared to the Bayes response and the upper bound \Cref{eq:oldupperbound}.}
    \label{fig:vsps}
\end{figure}

\section{Comparison with Indirect Rate-Distortion}
The similarity of the setting with indirect rate-distortion raises two questions: Is our scheme just solving the indirect rate-distortion problem with channel $\Pzgx$ and distortion $\Lambda$? Relatedly, are the compressors specified by our scheme necessarily ``good'' for the indirect rate-distortion problem? In this section we show that the answer is ``no'' to both questions outside of special cases.
Adding a perception constraint to the indirect rate-distortion problem, the two problems coincide under slightly weaker conditions, but are still generally different. We illustrate this case by an example with a memoryless Gaussian source $\xs$ and an AWGN channel $\Pzgx$.

\subsection{Indirect Rate-Distortion}
We note that in the indirect formulation the compressor is directly optimized to minimize the denoising loss. Hence, the indirect rate-distortion curve serves as a lower bound for the trade-off achieved by the compression-based denoiser. The following \Cref{prop:rdopttest} establishes the condition under which this lower bound is tight.
\begin{prop}
    \label{prop:rdopttest}
    The infimizing distribution in \Cref{thm:thmfour}, i.e. $\thedist_{\Y^k|\Z^k} = \thedist_{\X^k|\Z^k}$, achieves a point on the indirect rate-distortion curve $\ird(L)$ if there exists some $c_1\in \reals, c_2:\zc \to \reals$ such that for all $\z, \y$
    \begin{equation}
        \rho(\z, \y) = c_1\distdist(\z, \y) + c_2(\z).
    \end{equation}
    The condition is also necessary in the case when all alphabets are finite.
\end{prop}
In view of \Cref{thm:thmfour}, this also implies that the behavior of compressors designed for the distortion \cref{eq:lossdef} will in general behave differently than compressors designed for the indirect rate-distortion problem.
The difference is illustrated in the following example.
\blue{The results of \cref{sec:meat} were proved for the finite alphabet case. We appeal to the obvious generalization of the statement to continuous alphabets below. Methods to generalize the results of \cite{weissman_empirical_2005} to continuous alphabets can be found in \cite{shamai_empirical_1997}, notably Section~V. For results for Gaussian sources under MSE in a similar setting as below, see \cite[Section~V]{kipnis_rate-distortion_2021}}.
\newcommand{\snr}{\gamma}
\begin{example}[Gaussian Source, AWGN Channel]
\label{ex:scalargauss}
    Let $\X\sim \normal(0, 1)$ and $N\sim \normal(0,1)$ be independent.
    Let $\Z = \sqrt{\snr} \X+N$.
    We have $\X|\Z=\z \sim \normal\left(\z\frac{\sqrt{\snr}}{1+\snr}, \frac{1}{1+\snr}\right)$, and $H\left(\X\given\Z\right) = \frac{1}{2}\log\left(2\pi e \frac{1}{1+\snr}\right)$.
    We consider the MSE loss $\Lambda(x,y) = (x-y)^2$. 

    For the compression-based denoiser, we have $\rho(\z, \y) = \frac{1}{2}\left(\left(\sqrt{\snr} \y - \z\right)^2 + \log(2\pi)\right)$.
    \blue{For additive Gaussian noise of variance $\sigma^2$, the compression-based denoiser will always operate at distortion level
    \begin{equation}
        \label{eq:awgn-distortion}
    h(\Z|\X) = \frac{1}{2} \log (2\pi e \sigma^2).
    \end{equation}
    }
    We have $\Pygz = \thedist_{\X|\Z}$, achieving loss
    \begin{align}
        \exof[\Z]{\exof[(U,V)\sim\thedist_{\X|\Z}]{(U-V)^2}} = 2\frac{1}{1+\snr}.
    \end{align}
    The rate is
    \begin{align}
        R = I(\X;\Z) = \frac{1}{2}\log(1+\gamma).
    \end{align}

    For the indirect rate-distortion setting, we have
    \begin{align}
        \distdist(\z,\y) &= \exof{(\X-\y)^2\given \Z=\z}\\
        &= \left(\exof{\X\given \Z=\z} - y\right)^2 + \varof{\X\given \Z=\z}\\
        &=  \left(\z\frac{\sqrt{\snr}}{1+\snr} - y\right)^2 + \frac{1}{1+\snr}
    \end{align}
    Then $\Y$ achieves the rate-distortion for compressing $\X' := \Z\frac{\sqrt{\snr}}{1+\snr}$ under MSE distortion constraint $L' = L-\frac{1}{1+\snr}$.
    We have $\varof{\X'} = \frac{\snr}{1+\snr}$, and, from the rate-distortion function of a Gaussian source \cite{shannon_coding_1959}, we achieve rate
    \begin{align}
        R = \left(\frac{1}{2}\logfrac{\snr}{(1+\snr)L - 1}\right)_+.
    \end{align}
    To compare with the compression-based denoiser, we set $L =2\frac{1}{1+\snr} $, which yields a rate of
    \begin{align}
        R = \left(\frac{1}{2}\log(\snr)\right)_+.
    \end{align}
    For $\snr\leq 1$ we have $\Y=0$ always.
    Otherwise
    \begin{align}
        \Z|\Y=\y &\sim \normal\left(\y\frac{1+\snr}{\sqrt{\snr}}, \frac{1+\snr}{\snr}\right)\\
        \Y|\Z=\z &\sim \normal\left(\z\frac{\sqrt{\snr}}{1+\snr} \frac{\snr-1}{\snr}, \frac{1}{\snr}\frac{\snr-1}{1+\snr}\right)\\
        \Y &\sim \normal\left(0, \frac{\snr-1}{1+\snr}\right).
    \end{align}
    
    Alternatively, if we set $R = \frac{1}{2} \log(1+\snr)$, the indirect rate-distortion scheme achieves loss
    \begin{align}
        L=\frac{1+2\snr}{(1+\snr)^2} < 2\frac{1}{1+\snr}.
    \end{align}
\end{example}
In the above example it is possible to achieve the indirect rate-distortion curve by scaling the output of the compression-based scheme by $\frac{\snr}{1+\snr}$.
Designing compressors without knowledge of the denoising task but allowing for arbitrary post-processing is studied by Kipnis et al in \cite{kipnis_indirect_2015,kipnis_rate-distortion_2021}.
Their setting differs in that the post-processing allows for full knowledge of the source distribution and denoising loss, considering $\inf_f \exof{\Lambda_n(\X^n, f(\Y^n)}$ instead of $\exof{\Lambda_n\left(X^n, Y^n(Z^n)\right)}$.
\subsection{Indirect Rate-Distortion-Perception}
Optimality for the rate-distortion problem is often impossible because, by design, the reconstructions must have the same distribution as the source yet the indirect rate-distortion curve achieving $\thedist_{\Y^k|\Z^k}$ can in general be very different.
Adding an additional constraint to rate-distortion that the reconstruction resemble the source in distribution has been recently studied in rate-distortion-perception theory \cite{chen_information_2024, blau_rethinking_2019}.
This motivates the following comparison of the rate and the denoising performance of the compression-based denoiser to the following indirect rate-distortion curve with perfect perception constraint.
\newcommand{\irdp}{R_{\thedist_\X}} 

\begin{definition}
    We define the indirect rate-distortion curve with perfect perception constraint, denoted $\irdp(L)$, to be the solution to the optimization problem
    \begin{align}
        \min_{\Pygz} &I(\Z;\Y)\\
        \text{subject to} \quad &\exof[(\Y, \Z)\sim \Pygz \otimes \thedist_\Z]{\distdist(\Z, \Y)} \leq L\\
        & \Pygz \circ \thedist_\Z = \thedist_\X.
    \end{align}
    where
    \begin{equation}
        \distdist(\z,\y) := \exof{\Lambda(\X,\y)\given Z=z}.
    \end{equation}
\end{definition}
The following proposition characterizes the solution of this problem.
\begin{prop}
\label{prop:rdpopttest}
    \blue{Suppose $\xc, \yc, \zc$ are Polish spaces and equip them with their Borel $\sigma$-alebras (so that the resulting measurable spaces are standard Borel).}
    Suppose there exists a $\Pygz^*$ satisfying
    \begin{align}
        \label{eq:rdploss}
        \exof[\Pygz^*\otimes\thedist_\Z]{\distdist(\Z, \Y)}  &= L\\
        \Pygz^* \circ \thedist_\Z &= \thedist_\X\\
        \label{eq:asssume-abscont}
        \blue{\thedist_{Y|Z=z}^*} &\blue{\ll \thedist_\X \quad \thedist_{\Z}\text{-a.s.}}\\
        \label{eq:rdpexpcond}
        \frac{d\Pygz^*}{d\thedist_{\X}}(\y, \z) &= \exp\left(-\beta \distdist(z,y) + A(\y) + B(\z)\right)
    \end{align}
    for some $\beta, A, B, L$.
    \blue{Here, $\frac{d\Pygz^*}{d\thedist_{\X}}(\y, \z) = \frac{d\thedist_{Y|Z=z}^*}{dQ_{Y|Z=z}}(y)$, where $Q$ is the Markov kernel where $Y\sim P_X$ regardless of the value of $z$. From the standard Borel assumption and \cref{eq:asssume-abscont} we can conclude the right hand side is meaningful and jointly measurable \cite[Theorem~V.4.44]{cinlar_probability_2011}}.
    Then $\Pygz^*$ uniquely (up to $\thedist_\Z$-a.s.-equivalence) achieves $\irdp(L)$.
    The existence of such $\beta, A, B, L$ is also necessary in the case where all alphabets are finite.
\end{prop}

We next apply this proposition to the Gaussian case.
\blue{Again, as in \cref{ex:scalargauss}, we appeal to continuous analogs of the results of \cref{sec:meat}.}
\begin{example}[Gaussian Source, AWGN Channel]
    Let $\X_i\sim \normal(0, 1)$ and $N_i\sim \normal(0,1)$ be i.i.d.
    Let $\Z_i = \sqrt{\snr} \X_i+N_i$.
    We consider the MSE loss $\Lambda(x,y) = (x-y)^2$.
    Note that
    \begin{align}
        \distdist(\z,\y) &= \exof{(\X-\y)^2\given \Z=\z}\\
        &= \left(\exof{\X\given \Z=\z} - y\right)^2 + \varof{\X\given \Z=\z}\\
        &=  \left(\z\frac{\sqrt{\snr}}{1+\snr} - y\right)^2 + \frac{1}{1+\snr}.
    \end{align}
    By construction, setting $\Pygz^* = \thedist_{\X|\Z}$ satisfies the constraint $\Pygz \circ \thedist_\Z = \thedist_\X$.
    By Bayes' rule (and constraint that $\Y\disteq \X$) we have
    \begin{align}
        \frac{d\Pygz^*}{d\thedist_{\X}}(\y, \z) = \frac{d\Pzgy^*}{d\thedist_{\Z}}(\z, \y).
    \end{align}
    Then
    \begin{align}
         -\log\frac{d\Pygz^*}{d\thedist_{\X}}(\y, \z)
         &= \frac{1}{2}\snr \y^2 - \sqrt{\snr}\y\z + \frac{1}{2}\z^2 + \log\thedensity_\Z(\z)
    \end{align}
    We note  that only the $\sqrt{\snr}\y\z$ term depends on both $\z$ and $\y$.
    Similarly, the only term of $\distdist(\z,\y)$ that depends on both $\z$ and $\y$ is proportional to $\y\z$.
    Then \cref{eq:rdpexpcond} holds.
    We conclude by \Cref{prop:rdpopttest} that $\Pygz^*$ achieves $\irdp(L)$ for some $L$.

    For the compression-based denoiser choosing $\rho(\z, \y) = \frac{1}{2}\left(\left(\sqrt{\snr} \y - \z\right)^2 + \log(2\pi)\right)$ gives $\Pygz = \thedist_{\X|\Z}$, achieves loss
    \begin{align}
    \label{eq:scalarcompressloss}
        \exof[\Z]{\exof[(U,V)\sim\thedist_{\X|\Z}]{(U-V)^2}} = 2\frac{1}{1+\snr}
    \end{align}
and rate
    \begin{align}
    \label{eq:scalarcompressrate}
        R = I(\X;\Z) = \frac{1}{2}\log(1+\gamma).
    \end{align}
Evaluating \cref{eq:rdploss} and $I(\X;\Z)$ at the optimal solution $\Pygz^*$, we see that they match \cref{eq:scalarcompressloss} and \cref{eq:scalarcompressrate}.
We conclude the denoiser achieves $\irdp(L)$. This shows that in the scalar Gaussian case the compression based denoiser is able to achieve the optimal rate-distortion performance with perfect perception.
\end{example}

{
\section{Image Experiments with Off-the-Shelf Compressors (JPEG and neural)} \label{sec:experiment}
    \begin{figure}
    \centering
    \includegraphics[width=.7\linewidth]{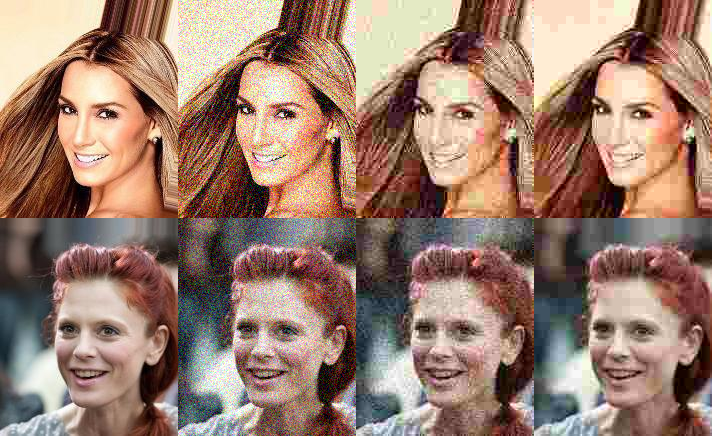}
    \caption{An example of images with Gaussian noise denoised using JPEG compression. From left to right, each row consists of a clean image, the image with additive Gaussian noise applied, the denoised image, and the clean image compressed to achieve the same rate as used for denoising.
    We have SNR $\gamma = 4$ ($\approx 3$ dB).}
    \label{fig:jpeg}
\end{figure}

\begin{figure}
    \centering
    \includegraphics[width=.7\linewidth]{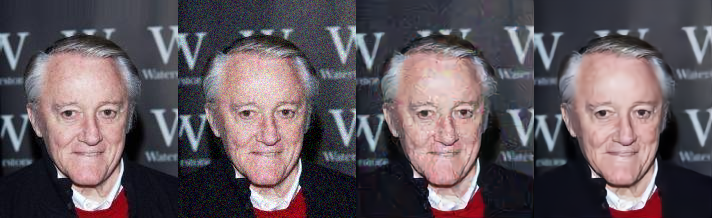}
    \includegraphics[width=.7\linewidth]{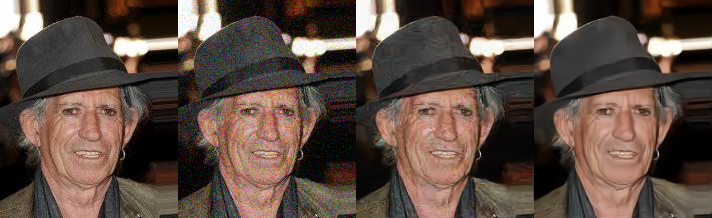}
    \includegraphics[width=.7\linewidth]{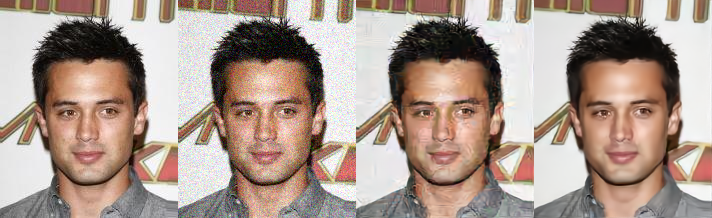}
    \caption{An example of images with Gaussian noise denoised using Efficient single-model Variable-bit-rate Codec \cite{guo-hua2023evc} as a compressor. From left to right, each row consists of a clean image, the image with additive Gaussian noise applied, the denoised image, and the clean image compressed to achieve the same rate as used for denoising.
    We have an SNR of 15 dB.}
    \label{fig:evc}
\end{figure}

\begin{figure}
    \centering
    \includegraphics[width=0.5\linewidth]{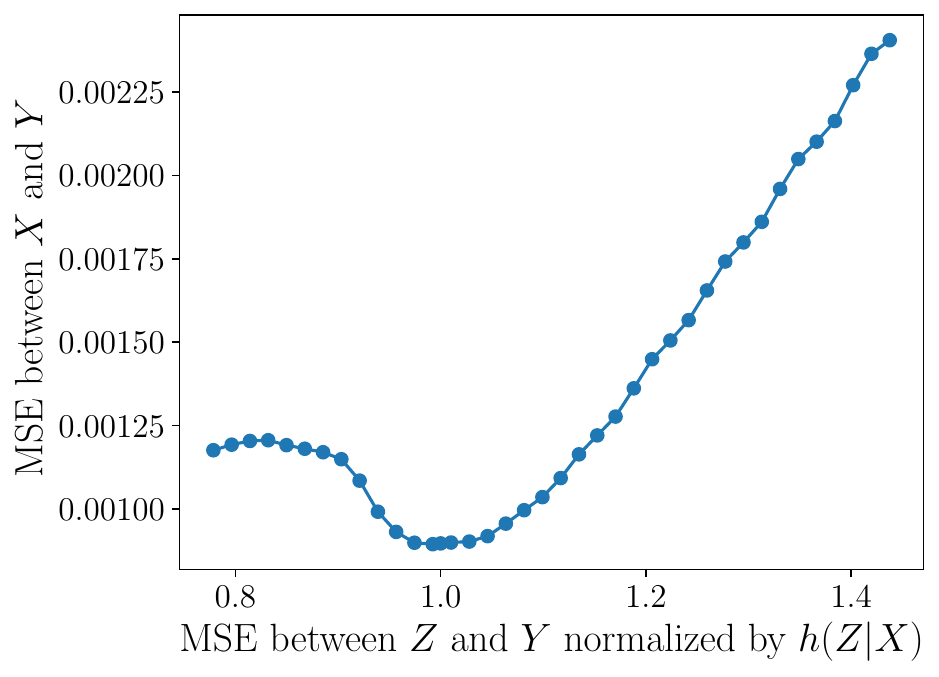}
    \caption{Plot of MSE between the reconstructed image $Y$ and the original image $X$ for various choices of the distortion level for the compressor using the image from the first row of \cref{fig:evc}. The $x$-axis represents the ratio of the distortion level and $h(Z|X)$. Values are averaged over 200 samples of the additive noise for each choice of distortion level.}
    \label{fig:level}
\end{figure}

The high-level conclusion from our theoretical framework in the previous sections is that compression can serve as a denoiser when the distortion metric and distortion level are chosen to match the channel characteristics. In this section, we test this proposition on noisy images using off-the-shelf compressors, including JPEG and neural compressors.

Although the sequence of pixels, or blocks of pixels, in an image can be viewed as a process, images do not strictly satisfy the stationarity assumptions of our theory. Similarly, off-the-shelf compressors are not necessarily optimal from a rate-distortion perspective, as they often involve heuristic optimizations and may even be treated as black boxes. Therefore, the experiments below should not be viewed as a strict validation of our theorems. Rather, they are intended to test whether the high-level principle of denoising via compression with a matched distortion criterion can be observed, at least to some extent, in a setting that substantially stretches the assumptions of our theory.

In the following, we use images from the CelebA dataset \cite{ziwei_liu_deep_2015}. Each color image is denoted by $\X \in [0,1]^{L\times W\times 3}$, where $L\times W$ is the image size and the last dimension corresponds to the three color channels. We corrupt each image $\X$ to obtain
$Z=X+N,$
where $N$ consists of pixel-wise and color channel-wise independent Gaussian noise with zero mean and variance $\sigma^2$.
    
As an approximation to a sequence of good codes, we apply off-the-shelf compressors with a distortion level chosen to match the channel. In the Gaussian noise case considered here, this corresponds to MSE between the compressor input and its reconstruction given by
$
h(Z|\X)=\frac{1}{2}\log(2\pi e\sigma^2),
$
as discussed in \cref{eq:awgn-distortion}.

\Cref{fig:jpeg} shows the results of compression-based denoising using JPEG, where we use the JPEG quality factor to control the distortion level. Similarly, \Cref{fig:evc} shows the denoising results obtained using a pretrained implementation\footnote{Checkpoint EVC\_SS\_MD from \url{https://github.com/microsoft/DCVC/blob/main/DCVC-family/EVC}.} of the Efficient single-model Variable-bit-rate Codec (EVC) \cite{guo-hua2023evc}, optimized for MSE, as the compressor. No modifications to the model were necessary for denoising. This neural compressor includes a scalar parameter that controls the number of quantization levels used in the quantization step, which we use to tune the distortion level.
    
In these experiments, each row shows, from left to right, the clean image, the noisy image obtained by adding pixel-wise Gaussian noise, the denoised image obtained by applying the compressor to the noisy image, and the compressed image obtained by applying the same compressor to the clean image. In the last column, the rate is adjusted to match the size of the compressed noisy image in the third column; thus, the third and fourth columns are matched in terms of rate, or equivalently, the number of bits used for the representation. 

Comparing the images in the second and third columns, we observe that compression indeed has a denoising effect. Furthermore, in \cref{fig:evc}, even when comparing the third and fourth columns, the reconstruction in the third column appears to have some perceptual advantages over the reconstruction in the fourth column. For example, in the first row of \cref{fig:evc} although the subject's skin appears smoother in the last column, the third column arguably provides a more realistic reconstruction that is more consistent with the skin texture in the original image.

We attribute this effect to the perfect-perception property of compression-based denoising, which manifests itself as increased realism. We note, however, that while the third and fourth columns are matched in rate, the comparison is inherently favorable to the fourth column: in the fourth column, the compressor is applied directly to the clean image, which also has lower entropy than the noisy image. Thus, the fact that the third column can compete perceptually with the fourth is somewhat surprising. In a sense, this suggests that, perceptually, it can sometimes be preferable to compress after adding noise. Of course, this would be fundamentally suboptimal from a classical distortion-only perspective. The advantage here arises because the compression-based denoising approach inherits perfect perception as a consequence of the framework, whereas the neural compressor itself may not be necessarily optimized for perceptual fidelity. This effect appears to be less pronounced for JPEG in \cref{fig:jpeg}.

In \cref{fig:level}, we empirically show that the reconstruction produced by the compression-based denoiser is closest to the ground truth, in terms of MSE, when the distortion level is chosen to match $h(Z|X)$, as suggested by our theoretical framework. The $y$-axis here denotes the MSE between the reconstructed image and the clean image, i.e., the MSE between $\Y$ and $\X$. The $x$-axis corresponds to the distortion level used by the compressor, measured as the MSE between the reconstructed image and the noisy image at the input of the compressor, i.e., between $\Y$ and $\Z$, normalized by $h(Z|X)=\frac{1}{2}\log(2\pi e\sigma^2).$
We observe that the reconstruction closest to the original image is indeed obtained when this ratio is equal to one. This provides empirical support for choosing the compression distortion to match the noise level of the channel in order to maximize the denoising effect, even in this setting, which substantially stretches the assumptions of our theory.

}

\section{Conclusion and Future Work} \label{sec:conclusion}
    In this paper we have established that lossy compression performs denoising for any stationary ergodic source observed through a DMC by outputting a sample from the posterior.
This was done by designing the distortion measure $\loss(\z,\y) = -\log\pzgx\left(\z\given \y\right)$ to match the channel and operating at a distortion level $D = H\left(\Z\given \X\right)$.
A key technical contribution was showing that,
under a mixing condition, the empirical distributions of the source $\X$ and the output $\Y$ given the observation $\Z$ approach conditional independence.
This lead to an exact \blue{asymptotic} expression for the loss achieved by the compression based denoiser as the expected loss of two independent samples from the posterior.
The substantial improvement of the characterization over previous bounds is demonstrated in several special cases.
Notably, when measuring denoising performance with MSE, the conditional independence results in a factor of 2 improvement over the previous bound.

Several directions remain for future work.
First, the results can be extended to almost-sure convergence and for general alphabets.
Second, characterizing the behavior of the denoiser operating at distortions $D\neq H\left(\Z\given \X\right)$ would provide insight to the tradeoffs available in the given framework, and studying the rate-distortion problem with distortion measure $\loss(\z,\y) = -\log\pzgx\left(\z\given \y\right)$ and arbitrary distortion level can be of independent interest.
Finally, additional experimental work applying the denoiser described here to real-world data would validate the utility of the given framework.
\section*{Acknowledgment}
This work was partially supported by the NSF grant CIF2213223. This work was partially supported by Stanford's SystemX Alliance. The authors would like to thank an anonymous reviewer for pointing out \cref{cor:metric} and \cref{rem:reviewer}.


%
\bibliographystyle{IEEEtran}
\bibliography{refs}
%
\appendices
\section{Deferred Proofs} \label{sec:proofs}
    \begin{IEEEproof}[Proof of \Cref{prop:markovmixing}]
    \newcommand{\exprate}{c}
    \newcommand{\badset}{A}
    \newcommand{\fudgefactor}{C_1}
    \newcommand{\minprob}{p_{\min}}
    It suffices to show the result for one-sided processes, as we can apply the one-sided result to $\tilde \X_i = (\X_{+i}, \X_{-i})$ for the two-sided result.
    Let $\minprob = \min_{\z, \x}\Pzgx\left(\z\given \x\right)$.
    By assumption $\minprob >0$.

    By assumption that $\xs$ is Markov, the setting reduces to a hidden Markov model.
    It is known that the initial hidden state is ``forgotten'' exponentially quickly in hidden Markov models.
    The following is a specialization of \cite[Theorem 2.2]{le_gland_exponential_2000}, see also \cite{ephraim_hidden_2002}, to deterministic initial distributions:
    \begin{lemma}[Exponential Forgetting]
    For all $\x, \x'$,
    \begin{equation}
        \begin{split}
            \limsup_{k\to\infty} \frac{1}{k} \log &\big\|\pof{\X_{k+1} = \cdot \given \Z^k, \X_0 = \x} \\
            &- \pof{\X_{k+1} = \cdot \given \Z^k, \X_0 = \x'}\big\|_{\mathrm{TV}} < 0
        \end{split}
    \end{equation}
    holds a.s., over the randomness of $\zs$.
    \end{lemma}
    By the data processing inequality, it follows that
    \begin{equation}
    \begin{split}
        &\limsup_{k\to\infty} \frac{1}{k} \log \biggr\|\pof{\Z_{k+1} = \cdot \given \Z^k, \X_0 = \x} \\
        &- \pof{\Z_{k+1} = \cdot \given \Z^k, \X_0 = \x'}\biggl\|_{\mathrm{TV}} < 0.
    \end{split}
    \end{equation}
    By equivalence of norms, we have for all $k$,
    \begin{equation}
        \begin{split}
            &\big\|\pof{\Z_{k+1} = \cdot \given \Z^k, \X_0 = \x} \\
            &- \pof{\Z_{k+1} = \cdot \given \Z^k, \X_0 = \x'} \big\|_{\infty} \leq \fudgefactor\exp(-\exprate k)
        \end{split}
    \end{equation}
    for some constants $\fudgefactor, \exprate >0$.

    Now we fix $\x_0$ and $\z^\infty$. 
    By Bayes' rule,
    \begin{align}
        &\thedist_{\X_0|\Z^{k+1}}\left(\x_0\given \z^{k+1}\right)\\
        &= \thedist_{\X_0|\Z^k}\left(\x_0\given \z^k\right) \frac{\thedist_{\Z_{k+1}|\X_0, \Z^k}\left(\z_{k+1}\given \x_0, \z^k\right)}{\thedist_{\Z_{k+1}|\Z^k}\left(\z_{k+1}\given \z^k\right)}\\
        \begin{split}
            &= \thedist_{\X_0|\Z^k}\left(\x_0\given \z^k\right)\thedist_{\Z_{k+1}|\X_0, \Z^k}\left(\z_{k+1}\given \x_0, \z^k\right)\\
            &\quad \exof{\thedist_{\Z_{k+1}|\X_0, \Z^k}\left(\z_{k+1}\given \X_0, \z^k\right)}\inv.
        \end{split}
    \end{align}
    We now show that the conditional distribution $\thedist_{\X_0|\Z^k}$ changes exponentially little as $k$ is increased.
    \begin{align}
        &\abs{\thedist_{\X_0|\Z^{k+1}}\left(\x_0\given \z^{k+1}\right) - \thedist_{\X_0|\Z^k}\left(\x_0\given \z^k\right)}\\
        &= \frac{\abs{\thedist_{\Z_{k+1}|\X_0, \Z^k}\left(\z_{k+1}\given \x_0, \z^k\right) - \thedist_{\Z_{k+1}|\Z^k}\left(\z_{k+1}\given \z^k\right)}} {\thedist_{\Z_{k+1}|\Z^k}\left(\z_{k+1}\given \z^k\right)}\\
        \begin{split}
        &\leq \minprob\inv \big|\thedist_{\Z_{k+1}|\X_0, \Z^k}\left(\z_{k+1}\given \x_0, \z^k\right) \\
        &\quad - \thedist_{\Z_{k+1}|\Z^k}\left(\z_{k+1}\given \z^k\right)\big|
        \end{split}\\
        \begin{split}
        &= \minprob\inv \big|\thedist_{\Z_{k+1}|\X_0, \Z^k}\left(\z_{k+1}\given \x_0, \z^k\right) \\
        &- \exof{\thedist_{\Z_{k+1}|\X_0, \Z^k}\left(\z_{k+1}\given \X_0, \z^k\right)\given \Z^k = \z^k}\big|
        \end{split}\\
        \begin{split}
        &\leq \minprob\inv \ex\big[\big|\thedist_{\Z_{k+1}|\X_0, \Z^k}\left(\z_{k+1}\given \x_0, \z^k\right) \\
        &\quad -\thedist_{\Z_{k+1}|\X_0, \Z^k}\left(\z_{k+1}\given \X_0, \z^k\right)\big| \Z^k = \z^k\big|\big]
        \end{split}\\
        &\leq \minprob\inv \exp\left(-\exprate k\right).
    \end{align}
    Finally, applying the triangle inequality
    \begin{align}
        &\abs{\thedist_{\X_0|\Z^k}\left(\x_0\given \z^k\right) - \thedist_{\X_0|\zs}\left(\x_0\given \z^{\infty}\right)}\\
        &\leq \sum_{k'\geq k}\abs{\thedist_{\X_0|\Z^{k'+1}}\left(\x_0\given \z^{k'+1}\right) - \thedist_{\X_0|\Z^{k'}}\left(\x_0\given \z^{k'}\right)}\\
        &\leq \minprob\inv\sum_{k'\geq k}\exp\left(-\exprate k\right),
    \end{align}
    which vanishes exponentially in $k$, as desired.
    Here we have implicitly used the fact that $\thedist_{\X_0|\Z_{-k}^k}\left(x_0\given \Z_{-k}^k\right) \toae \thedist_{\X_0|\Z_{-k}^k}\left(x_0\given \Z_{-\infty}^\infty \right)$.
\end{IEEEproof}

\begin{IEEEproof}[Proof of \Cref{lemma:mixingupperbound}]
    \newcommand{\localdiff}{D_i\left(Z^n\right)}
    Concretely, we want to show
    \begin{align}
    \label{eq:mixingLHS}
    \begin{split}
        \Big|&\exof{\empirical{\X^n, \Z^n, \Y^n}\left(\x_0, \z_{-k}^k, \y_{-k}^k\right)} - \\  &\thedist_{\X_0 | \Z_{-k}^k}\left(\x_0, \z_{-k}^k\right)\exof{\empirical{\Z^n, \Y^n}\left(\z_{-k}^k, \y_{-k}^k\right)}\Big|
    \end{split}\\
    &\quad\leq \mixcoeff_k \exof{\empirical{\Z^n, \Y^n}\left(\z_{-k}^k, \y_{-k}^k\right)}.
    \end{align}

    For all $i$, the following upper bound by $\mixcoeff_k$ holds $\thedist_{\Z^n}$-a.s.
    \begin{align}
        &\abs{\pof{\X_0 = \x_0 \given Z_{-k}^k} - \pof{\X_i = \x_0\given \Z^n}}\\
        &= \abs{\pof{\X_0 = \x_0 \given Z_{-k}^k} - \pof{\X_0 = \x_0\given \Z_{-i+1}^{n-i}}}\\
        &= \abs{\pof{\X_0 = \x_0 \given Z_{-k}^k} - \exof[\Z_{-\infty}^{-i}, \Z_{n-i+1}^{\infty}]{\pof{\X_0 = \x_0\given \Z_{-\infty}^{\infty}}}}\\
        &\leq \exof[\Z_{-\infty}^{-i}, \Z_{n-i+1}^{\infty}]{\abs{\pof{\X_0 = \x_0 \given Z_{-k}^k} - \pof{\X_0 = \x_0\given \Z_{-\infty}^{\infty}}}}\\
        &\leq \mixcoeff_k,
    \end{align}
    where the first equality uses stationarity and the first inequality uses Jensen's inequality.
    We denote 
    \begin{equation}
        \localdiff = \pof{\X_i = \x_0\given \Z^n} - \pof{\X_0 = \x_0 \given Z_{-k}^k}.
    \end{equation}
    We just proved above that $\abs{\localdiff}\leq \mixcoeff_k$ a.s.

    Factoring the indicator functions and conditioning on $\Z^n$, we can take advantage of the Markov structure $\X^n\mchain \Z^n \mchain \Y^n$,
    \begin{align}
        &\exof{\empirical{\X^n, \Z^n, \Y^n}\left(\x_0, \z_{-k}^k, \y_{-k}^k\right)}\\
        \begin{split}
        &= \frac{1}{n-2k}\sum_{i=k+1}^{n-k} \ex\bigl[\ind{\Z_{i-k}^{i+k} =\z_{-k}^k}\\
        &\exof{\ind{\X_i = x_0, \Y_{i-k}^{i+k} =\y_{-k}^k}\given \Z^n}\bigr]
        \end{split}\\
        \begin{split}
        &= \frac{1}{n-2k}\sum_{i=k+1}^{n-k} \ex\bigl[\ind{\Z_{i-k}^{i+k} =\z_{-k}^k}\\
        &\quad \exof{\ind{\X_i = x_0}\given \Z^n}\exof{\ind{\Y_{i-k}^{i+k} =\y_{-k}^k}\given \Z^n}\bigr]
        \end{split}\\
        \begin{split}
        &= \frac{1}{n-2k}\sum_{i=k+1}^{n-k} \ex\bigl[\ind{\Z_{i-k}^{i+k} =\z_{-k}^k}\\
        &\left(\pof{\X_0 = \x_0 \given Z_{-k}^k} + \localdiff \right)\exof{\ind{\Y_{i-k}^{i+k} =\y_{-k}^k}\given \Z^n}\bigr].
        \end{split}
    \end{align}
    Similarly, we can rewrite
    \begin{align}
        &\pof{\X_0 = \x_0 \given Z_{-k}^k=\z_{-k}^k}\exof{\empirical{\Z^n, \Y^n}\left(\z_{-k}^k, \y_{-k}^k\right)}\\
        \begin{split}
        &= \frac{1}{n-2k}\sum_{i=k+1}^{n-k}\pof{\X_0 = \x_0 \given Z_{-k}^k=\z_{-k}^k}\\
        &\exof{\ind{\Z_{i-k}^{i+k} =\z_{-k}^k}\exof{\ind{\Y_{i-k}^{i+k} =\y_{-k}^k}\given \Z^n}}
        \end{split}\\
        \begin{split}
        &= \frac{1}{n-2k}\sum_{i=k+1}^{n-k} \ex\biggl[\ind{\Z_{i-k}^{i+k} =\z_{-k}^k}\\
        &\qquad\qquad\pof{\X_0 = \x_0 \given Z_{-k}^k} \exof{\ind{\Y_{i-k}^{i+k} =\y_{-k}^k}\given \Z^n}\biggr].
        \end{split}
    \end{align}
    Subtracting the two previous displays yields
    \begin{align}
        \begin{split}
        &\biggl|\exof{\empirical{\X^n, \Z^n, \Y^n}\left(\x_0, \z_{-k}^k, \y_{-k}^k\right)} \\
        &- \thedist_{\X_0 | \Z_{-k}^k}\left(\x_0, \z_{-k}^k\right)\exof{\empirical{\Z^n, \Y^n}\left(\z_{-k}^k, \y_{-k}^k\right)}\biggr|
        \end{split}\\
        \begin{split}
        &= \biggl|\frac{1}{n-2k}\sum_{i=k+1}^{n-k} \ex\bigl[\ind{\Z_{i-k}^{i+k} =\z_{-k}^k}  \\
        &\qquad \localdiff \exof{\ind{\Y_{i-k}^{i+k} =\y_{-k}^k}\given \Z^n}\bigr]\biggr|
        \end{split}\\
        &\leq \mixcoeff_k\biggl|\frac{1}{n-2k}\sum_{i=k+1}^{n-k} \ex\bigl[\ind{\Z_{i-k}^{i+k} =\z_{-k}^k}\\
        &\qquad \exof{\ind{\Y_{i-k}^{i+k} =\y_{-k}^k}\given \Z^n}\bigr]\biggr|\\
        &= \mixcoeff_k \exof{\empirical{\Z^n, \Y^n}\left(\z_{-k}^k, \y_{-k}^k\right)},
    \end{align}
    as desired.
\end{IEEEproof}

\begin{IEEEproof}[Derivations for \Cref{ex:bsserasure}]
    We denote the closest time in the past, respectively future, that we observe a non-erased symbol as
    \begin{align}
        \tn &= \max\set{t\leq 0: \Z_t \neq \erased}\\
        \tp &= \min\set{t> 0: \Z_t \neq \erased}.
    \end{align}
    With this convention only $\tn$ can be $0$ and $\tp>\tn$ always.
    We use $-,+$ subscripts to denote the values of processes at $\tn, \tp$ respectively.
    By Markov property the observations at $\tn, \tp$ are all that is relevant for the posterior distribution of $\X_0$, i.e.
    \begin{equation}
        \thedist_{\X_0| \zs} = \thedist_{\X_0| \tp, \tn, \Z_{\tn}, \Z_{\tp}}.
    \end{equation}
    Concretely,
    \begin{align}
        &\pof{\X_0=\x_0\given \zs = \z_{-\infty}^{\infty}}\\
        &= \frac{\pof{\X\atn=\z\atn, \X_0 = \x_0, \X\atp = \z\atp\given\tn, \tp }}{\pof{\X\atn=\z\atn,X\atp = \z\atp \given\tn, \tp }}
    \end{align}
    Using \cref{eq:binsourcetimeskip} we have
    \begin{align}
        &\pof{\X\atn=\z\atn, \X_0 = \x_0, \X\atp = \z\atp \given\tn, \tp }\\
        &= \frac{1}{8} \left(\negfactor{\x_0}{\z\atn}\q^{-\tn}+1\right) \left(\negfactor{\x_0}{\z\atp}\q^{\tp}+1\right)\\
        &\pof{\X\atn=\z\atn,X\atp = \z\atp \given\tn, \tp }\\
        &= \frac{1}{4}\left(\negfactor{\z\atn}{\z\atp} \q^{\tp-\tn} + 1\right)
    \end{align}

    With an explicit posterior, we can evaluate the loss achieved by Bayes-optimal and compression-based denoisers.
    First, for the Bayes optimal denoiser the loss can be simplified as
    \begin{align}
        &\exof[\tn, \tp]{\exof[\X\atn, \X\atp]{\min_\x \pof{\X_0=\x\given \tn, \tp, \X\atn, \X\atp}\given \tn, \tp}}\\
        \begin{split}
        &= \ex_{\tn, \tp}\biggl[\sum_{\z\atn, \z\atp} \pof{\X\atn=\z\atn,X\atp = \z\atp \given\tn, \tp }\\
        &\qquad\qquad\min_\x \pof{\X_0=\x\given \tn, \tp, \X\atn, \X\atp}\biggr]
        \end{split}\\
        \begin{split}
        &= \frac{1}{8}\ex_{\tn, \tp}\biggl[\sum_{\z\atn, \z\atp} \min_\x  \left(\negfactor{\x}{\z\atn}\q^{-\tn}+1\right) \\
        &\qquad\qquad\left(\negfactor{\x}{\z\atp}\q^{\tp}+1\right)\biggr]
        \end{split}
    \end{align}
    The Bayes-optimal denoiser will always output the value of the observation that is closer in time to $\X_0$.
    It is readily shown that the above minimization is achieved with the opposite choice of $\x$ (as it is the error probability).
    We therefore introduce the following notation.
    We let $\tc, \tf$ denote the closer and farther time, respectively, i.e.
    \begin{align}
        \tc &= \begin{cases}
            \tn & -\tn\leq \tp\\
            \tp &\text{otherwise}
        \end{cases}\\
        \tf &= \begin{cases}
            \tp & -\tn\leq \tp\\
            \tn &\text{otherwise}
        \end{cases}.
    \end{align}
    Again we use corresponding subscripts to denote values at times $\tc, \tf$.
    Taking $\x = 1-\z\atc$,
    \begin{align}
        &\sum_{\z\atn, \z\atp} \min_\x  \left(\negfactor{\x}{\z\atn}\q^{-\tn}+1\right) \left(\negfactor{\x}{\z\atp}\q^{\tp}+1\right)\\
        &=\sum_{\z\atc, \z\atf} \min_\x  \left(\negfactor{\x}{\z\atc}\q^{\abs{\tc}}+1\right) \left(\negfactor{\x}{\z\atf}\q^{\abs{\tf}}+1\right)\\
        &=\sum_{\z\atc, \z\atf} \left(-\q^{\abs{\tc}}+1\right) \left(-\negfactor{\z\atc}{\z\atf}\q^{\abs{\tf}}+1\right)\\
        &= 2\left(-\q^{\abs{\tc}}+1\right) \left(\left(\q^{\abs{\tf}}+1\right) + \left(-\q^{\abs{\tf}}+1\right)\right)\\
        &= 4 \left(-\q^{\abs{\tc}}+1\right)
    \end{align}
    Since the erasure channel is memoryless we have
    \begin{align}
        -\tn &\sim \geom(1-\pe)\\
        \tp-1 &\sim \geom(1-\pe)
    \end{align}
    and furthermore $\tn, \tp$ are independent.
    We have that $\abs{\tc}$ is a mixture of a point mass at $0$ and a geometric distribution beginning at $1$
    \begin{align}
        \pof{\abs{\tc}= 0} &= 1-\pe\\
        (\abs{\tc}-1)\big|(\abs{\tc}>0) &\sim \geom(1-\pe^2).
    \end{align}
    Then the Bayes-optimal loss reduces to
    \begin{align}
        &\frac{1}{2}\exof{\left(-\q^{\abs{\tc}}+1\right)}\\
        &= -\frac{1}{2}\left((1-\pe)\q^0 + \pe\exof[T\sim \geom(1-\pe^2)]{\q^{T+1}}\right) + \frac{1}{2}\\
        &= -\frac{1}{2}\left((1-\pe)+ \pe\q\frac{1-\pe^2}{1-\pe^2\q}\right) + \frac{1}{2}\\
        &= \frac{\pe \ps}{1-\pe^2\qlong}.
    \end{align}
    
    We proceed with the compression-based scheme.
    First, the conditional expectation
    \begin{align}
        &\exof{F( \thedist_{\X_0| \tp, \tn, \Z_{\tn}, \Z_{\tp}}) \given \tn, \tp}\\
        \begin{split}
            &= 2\sum_{\z\atn, \z\atp} \frac{1}{\pof{\X\atn=\z\atn,X\atp = \z\atp \given\tn, \tp }}
            \\&\qquad\prod_{x\in\{0,1\}}\pof{\X\atn=\z\atn, \X_0 = x, \X\atp = \z\atp\given\tn, \tp }
        \end{split}\\
        \begin{split}
            &= \frac{1}{8} \sum_{\z\atn, \z\atp} \left(\negfactor{\z\atn}{\z\atp} \q^{\tp-\tn} + 1\right)\inv\\
            &\qquad\left(1-\q^{-2\tn}\right) \left(1-\q^{2\tp}\right)\\
        \end{split}\\
        &= \frac{1}{2}\frac{\left(1-\q^{-2\tn}\right) \left(1-\q^{2\tp}\right)}{1-\left(\q^{-2\tn}\right)\left(\q^{2\tp}\right)}.
    \end{align}
    Then the denoising loss can be given by the infinite sum:
    \begin{align}
        &\exof{F\left(\thedist_{\X_0|\zs}\right)}\\
        &= \frac{1}{2}(1-\pe)^2\sum_{s,t\geq 0} \pe^s\pe^{t} \frac{\left(1-\q^{2(t+1)}\right) \left(1-\q^{2s}\right)}{1-\left(\q^{2(t+1)}\right)\left(\q^{2s}\right)}.
    \end{align}
\end{IEEEproof}
\begin{IEEEproof}[Proof of \Cref{prop:rdopttest}]
    We use Theorem~9.4.1 (see also comments at the end of Section~9.6) from \cite{gallager_information_1968}.
    \begin{lemma}
    \label{lemma:gallager}
    The distribution $\Pygz$ minimizes $I(\Z; \Y)$ subject to $\exof{\distdist(\Z, \Y)}  \leq L$ if $\exof{\distdist(\Z, \Y)}  = L$ and the backward channel $\Pzgy$ satisfies
    \begin{equation}
        \label{eq:rdopttestexp}
        \pzgy\left(\z\given\y\right) = \exp\left(-\beta \distdist(\z, \y) + B(\z)\right)
    \end{equation}
    for some $B,\beta$ such that
    \begin{equation}
    \label{eq:rdopttestcond}
        \sum_\z  \exp\left(-\beta \distdist(\z, \y) + B(\z)\right) \leq 1
    \end{equation}
    for all $\y$ (i.e. including $\y$ outside of the support of $\thedist_\Y$).
    The only if direction holds if all alphabets are finite.
    \end{lemma}
    By definition of $\rho$,
    \begin{equation}
        \pzgy\left(\z\given \y\right) = \pzgx\left(\z\given \y\right) = \exp\left(-\rho(\z,\y)\right).
    \end{equation}
    At the same time, \Cref{lemma:gallager} gives the sufficient (and necessary in case of finite alphabets) condition
    \begin{equation}
        \pzgy\left(\z\given\y\right) = \exp\left(-\beta \distdist(\z, \y) + B(\z)\right).
    \end{equation}
    Equating the exponents yields
    \begin{equation}
        \rho(\z,\y) = \beta \distdist(\z, \y) - B(\z),
    \end{equation}
    so we can choose $c_1 = \beta$ and $c_2 = -B$ to satisfy \cref{eq:rdopttestexp}.
    We can assume we always choose to use a version of $\pzgx$ such that $\sum_{\z}\pzgx \leq 1$ for all $\x$.
    Then \cref{eq:rdopttestexp} implies \cref{eq:rdopttestcond}.
    Finally, we choose $L = \exof{\distdist(\Z, \Y)}$.
    Then applying \Cref{prop:rdopttest} gives the desired result.
\end{IEEEproof}
\begin{IEEEproof}[Proof of \Cref{prop:rdpopttest}]
    In the finite alphabet case, the proof is nearly identical to that of \Cref{prop:rdopttest}. We proceed with giving the general alphabet result for just the sufficient condition.
    
    We note that the problem is equivalent to
    \begin{align}
        \min_{\Pygz} &\exof[\thedist_\Z]{\fdiv{\Pygz\left(\cdot\given \Z\right)}{\thedist_{\X}(\cdot)}}\\
        \text{subject to} \quad &\exof[\Pygz \otimes \thedist_\Z ]{\distdist(\Z, \Y)} \leq L\\
        & \Pygz \circ \thedist_\Z = \thedist_\X.
    \end{align}
    Suppose $\Pygz$ is feasible, i.e. 
    \begin{align}
        \exof[\Pygz \otimes \thedist_\Z ]{\distdist(\Z, \Y)} &\leq L\\
        \Pygz \circ \thedist_\Z &= \thedist_\X.
    \end{align}
    Then,
    \begin{align}
        &\exof{\fdiv{\Pygz\left(\cdot\given \Z\right)}{\thedist_{\X}}}\\
        \label{eq:chainrule-step}
        \begin{split}
            &= \exof[\thedist_\Z ]{\fdiv{\Pygz\left(\cdot\given \Z\right)}{\Pygz^*\left(\cdot\given \Z\right)}} \\
            &\quad+ \exof[\Pygz \otimes \thedist_\Z]{\log\left(\frac{d\Pygz^*}{d\thedist_{\X}} \left(\Y\given \Z\right)\right)}
        \end{split}\\
        &\geq \exof[\Pygz \otimes \thedist_\Z]{\log\left(\frac{d\Pygz^*}{d\thedist_{\X}} \left(\Y\given \Z\right)\right)}\\
        &= \exof[\Pygz \otimes \thedist_\Z]{-\beta \distdist(\Z, \Y) + A(\Y) + B(\Z)}\\
        &= - \beta \exof[\Pygz \otimes \thedist_\Z]{\distdist(\Z, \Y)} + \exof[\Y\sim\thedist_{\X}]{A(\Y)} + \exof[\thedist_{\Z}]{B(\Z)}\\
        &\geq -\beta L + \exof[\Y\sim \thedist_{\X}]{A(\Y)} + \exof[\thedist_{\Z}]{B(\Z)}
        \\
        &= \exof[Y]{\exof[\Pzgy^*]{-\beta \distdist(\Z, \Y) + A(\y) + B(\z)}}\\
        &= \exof{\fdiv{\Pzgy^*\left(\cdot\given \Y\right)}{\thedist_{\Z}}}.
    \end{align}
    \blue{In \cref{eq:chainrule-step} we have used the chain rule for Radon-Nikodym derivatives $\frac{d\Pygz}{d\thedist_{\X}} = \frac{d\Pygz^*}{d\thedist_{\X}} \frac{d\Pygz}{d\Pygz^*}$.}
    Additionally, the inequalities hold with equality if and only if $\thedist_\Z$-a.s. we have $\Pygz = \Pygz^*$.
    We conclude $\Pygz^*$ is the unique achiever of $\irdp(L)$.
\end{IEEEproof}
\end{document}